\documentclass[10pt,aps,pra,twocolumn,showpacs,superscriptaddress,nofootinbib]{revtex4-1}
\usepackage[english]{babel} 
\usepackage[usenames,dvipsnames]{color} %color management
\usepackage{graphicx} % enhanced support for graphics
\usepackage{bm} % Bold mathsymbols
\usepackage{amsmath} % improvements for mathformulas
\usepackage{amsthm} % improvements theorem-like enviouronment
\usepackage{amssymb} % extended symbols
\usepackage{times} % fontes adobe
\usepackage[pdftex,colorlinks=true,urlcolor= blue,linkcolor=Blue,citecolor=RedViolet]{hyperref} %link navigation
\begin{document}
\title{Coupled Harmonic Systems as Quantum Buses in Thermal Environments}
\author{F. Nicacio} \email{fernando.nicacio@ufabc.edu.br  } 
\author{F. L. Semi\~ao}
\affiliation{ Centro de Ci\^encias Naturais e Humanas, 
              Universidade Federal do ABC,   
              09210-170, Santo Andr\'e, S\~ao Paulo, Brazil }
\date{\today}
%%%%%%%%%%%%%%%%%%%%%%%%%%%%%%%%%%%%%%%%%%%%%%%%%%%%%%%%%%%%%%%%%%%%%%%%%%%%%%%%%%%%%%%%%
\begin{abstract}
\noindent In this work, we perform a careful study of a special arrangement of 
coupled systems that consists of two external harmonic oscillators weakly coupled 
to an arbitrary network (data bus) of strongly interacting oscillators. 
Our aim is to establish simple effective Hamiltonians and Liouvillians 
allowing an accurate description of the dynamics of the external oscillators regardless of  
the topology of the network. 
By simple we mean an effective description using just a few degrees of freedom. 
With the methodology developed here, we are able to treat general topologies and, 
under certain structural conditions, 
to also include the interaction with external environments. 
In order to illustrate the predictability of the simplified dynamics, we present 
a comparative study with the predictions of the numerically obtained exact description 
in the context of propagation of energy through the network. 
\end{abstract}
%%%%%%%%%%%%%%%%%%%%%%%%%%%%%%%%%%%%%%%%%%%%%%%%%%%%%%%%%%%%%%%%%%%%%%%%%%%%%%%%%%%%%%%%%
\maketitle
%%%%%%%%%%%%%%%%%%%%%%%%%%%%%%%%%%%%%%%%%%%%%%%%%%%%%%%%%%%%%%%%%%%%%%%%%%%%%%%%%%%%%%%%%
%%%%%%%%%%%%%%%%%%%%%%%%%%%%%%%%%%%%%%%%%%%%%%%%%%%%%%%%%%%%%%%%%%%%%%%%%%%%%%%%%%%%%%%%%
\section{ Introduction }\label{introd}  %%%%%%%%%%%%%%%%%%%%%%%%%%%%%%%%%%%%%%%%%%%%%%%%%
%%%%%%%%%%%%%%%%%%%%%%%%%%%%%%%%%%%%%%%%%%%%%%%%%%%%%%%%%%%%%%%%%%%%%%%%%%%%%%%%%%%%%%%%%
%%%%%%%%%%%%%%%%%%%%%%%%%%%%%%%%%%%%%%%%%%%%%%%%%%%%%%%%%%%%%%%%%%%%%%%%%%%%%%%%%%%%%%%%%
The manipulation of small scale systems is a key feature of quantum technologies 
and their quantum behavior is an incontestable mark of the success of quantum mechanics. 
Such control is an important tool to reveal the potential of quantum concepts 
from a practical point of view as well. 
In this context, 
one may mention interacting nanoelectromechanical systems \cite{buks,eisert} whose 
harmonic movement of the elements can be used to harness the power of continuous 
variables (position, momentum, etc) for quantum information purposes.
Another important example of a scalable system for exploration of quantum dynamics is 
the one consisting of trapped ions whose positions are coupled through dipole-dipole 
interactions \cite{bermudez}.  
In both cases, one can end up with a practical implementation of a network of 
interacting harmonic oscillators that are encompassed in the object of study 
of this paper.
Advances in experimental implementations of oscillator networks in the context of 
optomechanics \cite{lin, massel,shkarin} may also be mentioned as potential candidates 
for implementation of the general results discussed here. 

Due to their prominent role in physics in general, and in quantum technologies 
in particular, networks of coupled harmonic oscillators are a 
timely topic of interest. 
In the context of entanglement, for instance, the characterization of 
equilibrium states were studied in \cite{audenaert}. 
On the other hand, entanglement dynamics was the subject treated in \cite{plenio2004}. 
Concerning favorable conditions for entanglement propagation, 
in \cite{plenio2005}, a clever scheme of minimal adjustments of 
frequencies and coupling constants is developed, 
which enables highly efficient transfer of entanglement through a linear 
chain (first neighbor interactions) of coupled harmonic oscillators. 
Starting from a pure numerical study, 
they also found a simplified Hamiltonian model (no thermal baths) which allowed 
analytical progress in the understanding of the high efficiency. 
This was achieved by using the rationale of the rotating wave approximation (RWA), {\it i.e.}, 
the elimination of  fast oscillating terms in the interaction picture Hamiltonian when 
they do not contribute appreciably for the dynamics.
Our goal is to expand such an idea of frequency and coupling constants adjustments in 
many different directions, not explored in \cite{plenio2005}. 

We present here conditions for obtaining simplified reduced models for general 
configurations or topologies of quadratically coupled systems 
and, more importantly, we work within the formalism of open quantum systems, allowing us 
to include the presence of surrounding environments. 
The contribution here allows one to envision applications of our general results to the 
study of thermal properties
\cite{assadian,martinez,freitas,nicacio2015},  
non-classical properties \cite{leandro} 
or non-equilibrium thermodynamics \cite{oscillators_th} in harmonic systems; 
all of them examples of timely topics of research.

In this work, we show that the indirect or dynamical coupling between two distinct 
oscillators mediated by a general network, the latter with an arbitrary number of degrees
of freedom and topology, can be effectively described by a simplified model containing 
only a few degrees of freedom, provided the RWA rationale used in \cite{plenio2005} is
generalized.
The first step of the method comprises the diagonalization of the Hamiltonian 
of the network, which reveals its normal modes. 
It is here that the symplectic formalism is needed %
\cite{nicacio2010,kyoko,ozorio,littlejohn,gosson}. 
Using this tool, we can provide a case-independent diagonalization, {\it i.e.}, 
a general procedure without specification {\it{a priori}} of the topology of the network.
The following critical point is to careful understand how the external distinct 
oscillators couple to the normal modes of the network, and this is highly dependent 
on resonances between normal mode frequencies of the network and the natural frequencies 
of the external oscillators. 
When dealing here with topologies which are more complex than a linear chain 
treated in \cite{plenio2005}, we must take into account possible degeneracies in the 
frequencies of the normal modes in order to find the correct effective simplified model. 
Adding to that, another distinctive feature of our work is the inclusion of an environment 
in the dynamics. 
In this respect, we 
extend the unitary description in \cite{plenio2005} to a non-unitary open system 
treatment of the dynamics following a Lindblad master equation (LME). 
Obtaining an effective description in terms of just a few degrees of freedom when 
thermal baths are present is not a trivial task. However, under certain structural 
conditions, we were able to successfully perform such a simplification as discussed 
in this work. 
The designed methodology is suitable for the study of communication and transport across
the network and, as an illustration or application, we explore here the phenomenon of 
energy transfer between the external oscillators mediated by the network.    

The paper is organized as follows. 
In section \ref{tsid}, we describe the system of interest, namely two external 
oscillators coupled to a general network of oscillators. 
We add also the presence of thermal baths. 
Notation in the scope of the formalism of continuous variables and the dynamical 
equations for the open system dynamics are also presented in this section.  
The development of the method to obtain simplified models for the dynamics of the 
external oscillators is presented in Sec.~\ref{ed}. 
There, 
we present the mathematical tools of the symplectic 
formalism needed to perform the diagonalization of the network Hamiltonian 
and to obtain its normal modes. 
We also present conditions under which an RWA for the open system can be performed, 
which enable us to drastically simplify the system dynamics description to a picture 
with just a few degrees of freedom only.
Sec.~\ref{elc} is devoted to an example of the previous simplified description, 
a linear chain. 
In the context of quantum technologies and the usefulness of the simplified descriptions,
we will study the propagation of energy from a quantum system to another 
through a quantum bus in Sec.~\ref{et}, where we start with the case of 
a network (linear chain) with non-degenerate normal modes and then proceed to an 
interesting example with degeneracy. 
We end this section with a consideration of a network not obeying the Hooke's law. 
We then conclude with final remarks and some perspectives 
of future work in Sec.~{\ref{fr}}.
Additionally, two appendices are dedicated to details about the  RWA,  
which is a fundamental tool used in the work, and to long length analytic expressions.  
%%%%%%%%%%%%%%%%%%%%%%%%%%%%%%%%%%%%%%%%%%%%%%%%%%%%%%%%%%%%%%%%%%%%%%%%%%%%%%%%%%%%%%%%%
%%%%%%%%%%%%%%%%%%%%%%%%%%%%%%%%%%%%%%%%%%%%%%%%%%%%%%%%%%%%%%%%%%%%%%%%%%%%%%%%%%%%%%%%%
\section{ The System and Its Dynamics }\label{tsid}  %%%%%%%%%%%%%%%%%%%%%%%%%%%%%%%%%%%%
\begin{figure}[!hbt]                                                      
\includegraphics[width=8cm]{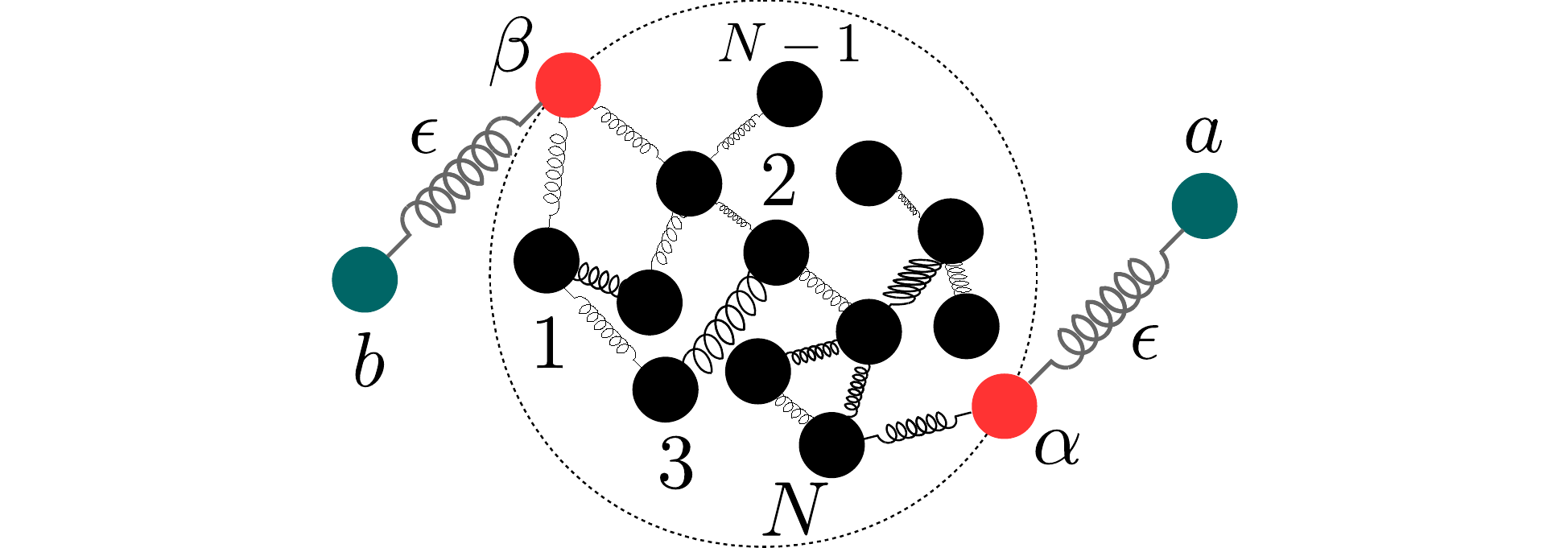} 
\caption{%
Schematic representation of the system of interest. 
It consists of a general network of $N$ oscillators 
where members  $\alpha^\text{th}$ and $\beta^\text{th}$  are 
coupled to two distinctive external oscillators denoted, 
respectively, as $a$ and $b$.  The coupling constant is 
$\epsilon$. At this level, we let the coupling constants inside 
the network completely arbitrary.}                                                       \label{fig1system1}                            
\end{figure}

%%%%%%%%%%%%%%%%%%%%%%%%%%%%%%%%%%%%%%%%%%%%%%%%%%%%%%%%%%%%%%%%%%%%%%%%%%%%%%%%%%%%%%%%%
%%%%%%%%%%%%%%%%%%%%%%%%%%%%%%%%%%%%%%%%%%%%%%%%%%%%%%%%%%%%%%%%%%%%%%%%%%%%%%%%%%%%%%%%%
The system we have in mind is depicted in  Fig.\ref{fig1system1}. 
Its temporal evolution will be governed by a LME 
for the density operator $\hat \rho$:
\begin{equation}                                                                         \label{lindblad}
\frac{d \hat \rho }{d t}  = 
         \frac{i}{\hbar}  [ \hat \rho , \hat H ] + \mathcal{L}(\hat\rho), 
\end{equation} 
where $\hat H$ is the Hamiltonian of the system and $\mathcal{L}(\hat\rho)$ is the 
non-unitary part of the dynamics accounting for the environment-system interaction. 
In the Lindblad scenario, the effect of the coupling between the system 
and the reservoir appears in Eq.~(\ref{lindblad}) by means of  
\begin{eqnarray}                                                                         \label{liouv}
\mathcal {L}(\hat \rho) = 
- \frac{1}{2\hbar} \! \sum_{k} \!\! 
             \left( 
                    \{ \hat{L}_{(k)}^\dag \hat{L}_{(k)} , \hat \rho  \}
                   - 2 \hat{L}_{(k)} \hat \rho \hat{L}_{(k)}^\dag   \right),                   
\end{eqnarray}                    
where the different $\hat{L}_{(k)}$ are known as the Lindblad operators, 
and $\{\hat A,\hat B\}$ denotes the anticommutator 
between general operators $\hat A$ and $\hat B$. 
The index $k$ of the sum is arbitrary until this point, 
it records the number of Lindblad operators needed to represent the reservoir 
interaction with the system.
%
%%%%%%%%%%%%%%%%%%%%%%%%%%%%%%%%%%%%%%%%%%%%%%%%%%%%%%%%%%%%%%%%%%%%%
%%%%%%%%%%%%%%%%%%%%%%%%%%%%%%%%%%%%%%%%%%%%%%%%%%%%%%%%%%%%%%%%%%%%% 

Let us define the collective operator  
$\hat X := ( \hat{\sf x}^\dagger, \hat x^\dagger)^\dagger$ as a column 
vector corresponding to the positions and the canonical conjugate 
momenta of the oscillators of the system. 
In this notation, and with respect to the system depicted in Fig.~\ref{fig1system1}, 
the operator 
$\hat x :=  (\hat q_1,...,\hat q_N, \hat p_1,...,\hat p_N)^\dagger $ 
accounts for the elements of the newtwork, while 
$\hat{\sf x} :=  
(\hat {\sf q}_a,\hat {\sf q}_b, 
 \hat {\sf p}_a,\hat {\sf p}_b )^\dagger $
represents the external oscillators.
The canonical commutation relations among coordinates and momenta 
are expressed compactly as 
\begin{equation}                                                                         \label{comm}
[\hat x_j , \hat x_k ] = i \hbar \, \mathsf J^{[N]}_{jk} ,  \,\,\, 
[\hat {\sf x}_j , \hat {\sf x}_k ] = i \hbar \, {\sf J}^{[2]}_{jk} ,  \,\,\, 
  \end{equation}  
where
\begin{equation}
\mathsf J^{[n]} := 
\left( \!\! \begin{array}{rc} 
       {\bf 0}_{n} & \mathsf I_n  \\
      -\mathsf I_n & {\bf 0}_n 
       \end{array}
\!\! \right)        
\end{equation}
is the fundamental $2n \times 2n$ symplectic matrix and the blocks 
$\mathsf I_n$ and ${\bf 0}_n$ are, respectively,  
the $n$ dimensional identity and zero matrices.
Even more compactly, one can write 
\begin{equation}                                                                         \label{comm2}
[\hat X_j , \hat X_k ] = i \hbar \, \mathsf J_{jk} 
\,\,\, \text {with} \,\,\, \mathsf J = \mathsf J^{[2]} \oplus \mathsf J^{[N]}.
\end{equation}  
The label $[n]$ of the above matrices is the number of degrees of freedom and will 
be omitted if clear in the context. 

The Hamiltonian $\hat H$ of the global system in (\ref{lindblad}) 
contains two parts $\hat H = \hat H_0 + \hat H_I$, 
where
\begin{equation}                                                                         \label{hamfree}
\hat H_0 = \tfrac{1}{2} \hat X^{\dagger} {\bf H}_0 \hat X  =  
\tfrac{1}{2} \hat{\sf x}^{\dagger} {\mathbf H}_{\rm e} \hat {\sf x} +
\tfrac{1}{2} \hat x^{\dagger} {\mathbf H}_{\rm N} \hat x  
\end{equation}
is the sum of the free Hamiltonians of network and external oscillators, 
and $\hat H_I$ describes the interaction of these subsystems. Assuming the 
standard Hooke's law prescription (springs), one can write   
\begin{equation}                                                                         \label{hamint} 
\hat H_I = \frac{\epsilon}{4} \left(\hat q_\alpha - \hat {\sf q}_a \right)^2 + 
           \frac{\epsilon}{4} \left(\hat q_\beta - \hat {\sf q}_b \right)^2.   
\end{equation}

As final remarks about the system Hamiltonian, note that in our notation, we have 
${\bf H}_0 = {\mathbf H}_{\rm e} \oplus {\mathbf H}_{\rm N}$. 
Finally, given the forms of (\ref{hamfree}) and (\ref{hamint}), 
the system Hamiltonian $\hat H$ is quadratic in $\hat X$, 
that is 
\begin{equation}                                                                         \label{hamtot}
\hat H = \tfrac{1}{2} \hat X^\dagger {\bf H} \hat X , 
\end{equation}
with $\bf H$ beeing the Hessian of $\hat H$. 
It is worth mentioning that, 
despite the specificity of the coupling Hamiltonian (\ref{hamint}), 
both ${\mathbf H}_{\rm e}$ and ${\mathbf H}_{\rm N}$ in (\ref{hamfree}) 
are completely arbitrary until this point.

For the non-unitary part of (\ref{lindblad}), let us assume that every 
$\hat{L}_{(k)}$ in (\ref{liouv}) is a linear function of position and momentum, 
{\it i.e.},  
\begin{equation}                                                                         \label{lindef}
\hat{L}_{(k)} =  \lambda_{(k)}^{\!\top} \mathsf J \hat{X}, 
\end{equation}
where $\lambda_{(k)} \in \mathbb C^{2N+4}$ is a column vector and 
$\mathsf J$ the matrix defined in (\ref{comm2}). 

Of particular importance for continuous variable systems, 
one defines the covariance matrix (CM) of the system state as
\begin{equation}                                                                         \label{cmdef}
\mathbf V_{\! jk} (t)  =  
\tfrac{1}{2} {\rm Tr}\left[ 
                           \left\{ \hat X_j - \langle \hat X_j \rangle_t , 
                           \hat X_k - \langle \hat X_k \rangle_t \right\}
                           \hat \rho(t)
                     \right],   
\end{equation}
where $\hat X_j $ is the $j^\text{th}$ component of $\hat X$ 
and
$\langle \hat X_j \rangle_t : = {\rm Tr}[ \hat X_j \hat \rho(t) ]$ 
is its mean value. 
Given its importance, we will be focusing on the time evolution of the CM in this paper.

With the help of (\ref{hamtot}) and (\ref{lindef}), and by defining
\begin{equation}                                                                         \label{decmatdef} 
{\bf \Upsilon} := \sum_{k } \lambda_{(k)} \lambda_{(k)}^\dagger, 
\end{equation}
it is possible to show that the CM equation of motion reads \cite{nicacio2010,kyoko} 
\begin{equation}                                                                         \label{cmev}
\frac{d}{d t} \mathbf V = 
{ \bf \Gamma }\mathbf V + \mathbf V {\bf \Gamma}^\top + {\bf D} , 
\end{equation}
with
\begin{eqnarray}                                                                         \label{dynmatdef}
{\bf \Gamma } := \mathsf J \mathbf H - {\rm Im} {\bf \Upsilon} \mathsf J , 
\,\,\,\, 
{\bf D} := \hbar \, {\rm Re}{\bf \Upsilon} .
\end{eqnarray}
Since $\bf H$ and $\bf \Upsilon$ are {\it time independent}, 
the solution for (\ref{cmev}) is
\begin{equation}                                                                         \label{cmsol}
{\bf V}(t)  =  {\rm e}^{{\bf \Gamma} t} \, {\bf V}\!_{0}  \, 
            { \rm e}^{ {\bf \Gamma}^{\! \top} t } +
 \int_0^t \! dt^\prime \, 
               {\rm e}^{{\bf \Gamma} t^\prime} \, 
                  {\bf D}  \,
               {\rm e}^{{\bf \Gamma}^{\! \top} t^\prime}  \, ,
\end{equation}
where ${\bf V}\!_0$ is the CM of the initial state. 

As a final remark, for initial Gaussian states, 
the qua\-dra\-tic Hamiltonians and linear Lindbladians considered here will 
dynamically preserve Gaussian states and, for this case, the CM and the mean values 
embody all information about the system. 
However, even in cases where the initial state is not Gaussian, 
(\ref{cmsol}) is still correct. 
In such cases, the knowledge of the CM and of the mean values  
will not contain all information about the system state.

%%%%%%%%%%%%%%%%%%%%%%%%%%%%%%%%%%%%%%%%%%%%%%%%%%%%%%%%%%%%%%%%%%%%%%%%%%%%%%%%%%%%%%%%%
%%%%%%%%%%%%%%%%%%%%%%%%%%%%%%%%%%%%%%%%%%%%%%%%%%%%%%%%%%%%%%%%%%%%%%%%%%%%%%%%%%%%%%%%%
\section{ Effective Dynamics }\label{ed}          %%%%%%%%%%%%%%%%%%%%%%%%%%%%%%%%%%%%%%%
%%%%%%%%%%%%%%%%%%%%%%%%%%%%%%%%%%%%%%%%%%%%%%%%%%%%%%%%%%%%%%%%%%%%%%%%%%%%%%%%%%%%%%%%%
%%%%%%%%%%%%%%%%%%%%%%%%%%%%%%%%%%%%%%%%%%%%%%%%%%%%%%%%%%%%%%%%%%%%%%%%%%%%%%%%%%%%%%%%%
In this section, we will expand and generalize the results in \cite{plenio2005} 
in order to treat arbitrary networks
possessing a quadratic and positive definite Hamiltonian. Besides, we include the 
non-unitary contribution to the dynamics via a Lindblad 
equation. This last point adds interest to our generalizations due to the multitude of 
physical systems where dissipation can not be neglected in the description.
The method consists in three steps. First, we 
diagonalize the free part of the system Hamiltonian. Then, we move the LME to an 
interaction picture where we can perform the RWA and a structural simplification 
to end up with an effective description for the dynamics of $a$, $b$, 
and a few normal modes of the network. 
%
%%%%%%%%%%%%%%%%%%%%%%%%%%%%%%%%%%%%%%%%%%%%%%%%%%%%%%%%%%%%%%%%%%%%%%%%%%%%%%%%%%%%%%%%%
\subsection{ Symplectic Formalism } \label{sf}
%%%%%%%%%%%%%%%%%%%%%%%%%%%%%%%%%%%%%%%%%%%%%%%%%%%%%%%%%%%%%%%%%%%%%%%%%%%%%%%%%%%%%%%%%
The development of our results is based on mathematical tools related to the symplectic 
formalism \cite{nicacio2010,kyoko,ozorio,littlejohn,gosson}. 
For the sake of simplicity, the basics will be illustrated using the $N$ 
oscillators of the network, but everything is readily transposed to 
systems with an arbitrary number of members. 

In this formalism, one is interested in transformations 
$\hat x' = \mathsf S \hat x$ of 
$\hat x=(\hat q_1,...,\hat q_N,\hat p_1,...,\hat p_N)^\dag$ 
such that the transformed operators satisfy  
\begin{equation}                                                                         \label{comm3}
[\hat x'_j , \hat x'_k ] = i \hbar \, \mathsf J^{[N]}_{jk}. 
\end{equation}
One can show that this is guaranteed provided  
$\mathsf S^{\!\top} \! \! \mathsf J \mathsf S = \mathsf J$.
The set of elements $\mathsf S$ satisfying such a statement forms the real 
symplectic group  $\mathsf S \in {\rm Sp}(2N, \mathbb R)$.

A central result for us here is the so called 
Williamson theorem \cite{williamson}. 
It states that a positive definite $2N \times 2N$ symmetric matrix $\mathbf M $, {\it i.e.}, 
$\mathbf M = \mathbf M^\top > 0$,  
can be diagonalized by a symplectic congruence. In other words, there exists 
$\mathsf S \in {\rm Sp}(2N, \mathbb R)$ such that 
\begin{equation}                                                                         \label{tw1}      
\mathsf S \mathbf M \mathsf S^\top 
= \Lambda_\mathbf{M}, \,\,\,  
\Lambda_\mathbf{M} := 
{\rm Diag}(s_1,...,s_N,s_1,...,s_N)  
\end{equation}
with
$ 0 < s_j \le s_k  \,\,\, \text{for} \,\,\, j \le k. $
The double-paired ordered set (or the diagonal matrix) $\Lambda_\mathbf{M}$ 
is called {\it symplectic spectrum} of $\mathbf M$, and $s_k$ are 
its symplectic eigenvalues (SE). 
These can also be found from the (Euclidean) eigenvalues of 
$\mathsf J \mathbf M$ \cite{gosson}, which turn out to be 
\begin{equation}                                                                         \label{tw2}
{\rm Spec_{\mathbb C}}(\mathsf J \mathbf M) = 
{\rm Diag}(is_1, ...,is_N,-is_1,...,-is_N).
\end{equation}
The matrix $\mathsf S$ that diagonalizes $\bf M$ admits a suitable decomposition as
\begin{equation}                                                                         \label{tw2a}
\mathsf S = \Lambda_{\bf M}^{\tfrac{1}{2}}  O \mathbf{M}^{-\tfrac{1}{2}} 
\end{equation}
with $O \in {\rm O}(2N)$, {\it i.e.}, an orthogonal matrix. 
From the symplectic condition on $\mathsf S$, one can see that $O$ must obey 
\begin{equation}\label{must}
O{\bf M}^{\tfrac{1}{2}} \mathsf J {\bf M}^{\tfrac{1}{2}} O^\top = 
\Lambda_{\bf M} \mathsf J.
\end{equation}

If convenient, one can equivalently use creation/anni\-hi\-la\-tion 
operators instead of position and momentum. In this case, 
one can define the column vector 
\begin{equation}                                                                         \label{zdef}
\hat z := (\hat a_1^\dagger,..., \hat a_N^\dagger, 
         - i \hat a_1,..., -i \hat a_N )^\dagger  , 
\end{equation}
where $\hat a_k := (\hat q_k + i \hat p_k)/\sqrt{2\hbar}$ 
is the annihilation operator. 
This change of coordinates can be compactly represented by 
\begin{equation}                                                                         \label{carep}
\sqrt{\hbar} \, \hat z = {\mathsf C}_{[N]} \hat x 
\end{equation}
with
\begin{equation}                                                                         \label{ctrans}
{\mathsf C}_{[n]} := \frac{1}{\sqrt{2}}
\left(\begin{array}{cc}
       \mathsf I_n   & i \mathsf I_n \\
       i \mathsf I_n & \mathsf I_n      
      \end{array} \right) ,
\,\,\, {\mathsf C}_{[n]}^\dag = {\mathsf C}_{[n]}^{-1}.  
\end{equation}
One can show that $\mathsf C_{[n]} $ is symplectic, and this leads immediately to 
\begin{equation}                                                                         \label{cpcr}
[\hat z_j , \hat z_k ] = i \, \mathsf J^{[N]}_{jk}.
\end{equation}  
If clear in context, the sub- or superscript $[n]$ will be omitted.   

Giving $\mathsf S$ as defined in (\ref{tw2a}), one may wonder which conditions should 
be imposed to $\bf M$ in order to make $\mathsf S \mathsf S{\!^\top}$ diagonal.  
To answer this question, let us define the diagonal matrix 
$\mathsf L := {\pmb L} \oplus {\pmb L}^{-1} \in {\rm Sp}(2N,\mathbb R)$, where 
${\pmb L}:= {\rm Diag}(l_1,...,l_N)$ with $l_i > 0 \forall i$. 
According to {\it theorem 5} in \cite{simon}, there exists a symplectic rotation 
$\mathsf R \in {{\rm Sp}(2N,\mathbb R) \cap {\rm O}(2N)} $ such that 
\begin{equation}                                                                         \label{theo5}
\mathsf R {\bf M} \mathsf R^\top = \mathsf L \, \Lambda_{\bf M}\,  \mathsf L^{\!\top}, 
\end{equation}
if and only if 
\begin{equation}                                                                         \label{condtheo}
\begin{aligned}
&[ {\bf M_Q},{\bf M_P} ] = {\bf M_C}^{\!\!2} - {{\bf M}_{\bf C}^{\top}}^2,  \\
&{\bf M_P} {\bf M_C} - {\bf M_C^\top} {\bf M_P} = 
 {\bf M_C}{\bf M_Q} - {\bf M_Q}{\bf M_C^\top}, 
\end{aligned}
\end{equation}
where we wrote $\bf M$ in terms of $N \times N$ blocks, {\it i.e.}, 
\begin{equation}
\bf M = \left( 
              \begin{array}{lc} 
              \bf M_Q      & \bf M_C \\ 
              \bf M_C^\top & \bf M_P 
              \end{array}               \right).  
\end{equation}

Once conditions (\ref{condtheo}) are fulfilled, 
the choice $\mathsf S := \mathsf L^{-1} \mathsf R$ 
leads to  $\mathsf S \mathbf M \mathsf S^\top = \Lambda_{\bf M}$, 
as a direct consequence of (\ref{theo5}). In this way, 
\begin{equation}
\mathsf S \mathsf S^\top = \mathsf L^{-1} \mathsf R (\mathsf L^{-1} \mathsf R)^\top = 
\mathsf L^{-2} = {\pmb L}^{-2} \oplus {\pmb L}^{2}  
\end{equation}
which is a diagonal matrix giving the way $\pmb L$ was defined.

%%%%%%%%%%%%%%%%%%%%%%%%%%%%%%%%%%%%%%%%%%%%%%%%%%%%%%%%%%%%%%%%%%%%%%%%%%%%%%%%%%%%%%%%%
\subsection{ Diagonalization of \texorpdfstring{$\hat H_0$}{H0}} \label{di}
%%%%%%%%%%%%%%%%%%%%%%%%%%%%%%%%%%%%%%%%%%%%%%%%%%%%%%%%%%%%%%%%%%%%%%%%%%%%%%%%%%%%%%%%%
We now require the matrix ${\bf H}_0$ appearing in (\ref{hamfree}) 
to be positive definite, and this is the only restriction imposed on the 
network of Fig.~\ref{fig1system1}. 
On the basis of the Williamson theorem, we know that there is a symplectic matrix
$\mathsf S_{0} = {\mathsf S}_{\rm e} \oplus {\mathsf S}_{\rm N}$ such that 
\begin{equation}                                                                         \label{hamtw}
\mathsf S_{0} {\bf H}_0 \mathsf S_{0}^\top 
= \Lambda_\mathrm{e} \oplus \Lambda_\mathrm{N},     
\end{equation}
where 
\begin{equation}                                                                         \label{nettw1}   
\Lambda_\mathrm{N} := {\rm Diag}(\varsigma_1,...,\varsigma_N,
                                 \varsigma_1,...,\varsigma_N)
\end{equation}
is the symplectic spectrum of $\bf H_{\rm N}$.

For the external oscillators, 
we may particularize to the case where they are identical with natural frequencies
$\Omega$ and masses $M$. In this case, their contribution to the Hamiltonian in 
(\ref{hamfree}) is given by
\begin{equation}                                                                         \label{hamext}
{\bf H}_{\rm e} =   M\Omega^2 \, \mathsf I_{2}  \oplus M^{-1} \mathsf I_{2}, 
\end{equation}
with symplectic spectrum $\Lambda_\mathrm{e} = \Omega \mathsf I_4$. 
One interesting situation arises when  ``$M\Omega = 1$" in a given system of units, 
for example, kilogram times radian. 
One can see that, in this case, Hamiltonian (\ref{hamext}) 
will be directly expressed 
in terms of normal mode coordinates with doubly degenerate frequency $\Omega$. 

The normal modes $\hat Y := (\hat{\sf y}^\dag, \hat{y}^\dag)^\dag$ for the whole system,  
by definition, relates to the original coordinates through  
\begin{equation}                                                                         \label{sdt}
\hat X = \mathsf S_0^{\top} \! \hat Y =  
\left(
\begin{array}{c}
{\mathsf S}_{\rm e}^\top \hat{\sf y} \\
{\mathsf S}_{\rm N}^\top \hat{y}
\end{array}
\right) ,
\end{equation}
where $\mathsf S_0$ is the symplectic transformation that diagonalizes 
the Hessian ${\bf H}_0$ in (\ref{hamtw}).  
In terms of creation/annihilation operators,  
\begin{equation}                                                                         \label{nmcoord}
\sqrt{\hbar} \, \hat Z = 
\left(  {\mathsf C}_{[2]} \oplus {\mathsf C}_{[N]} \right) \hat Y,     
\end{equation} 
which implies by (\ref{sdt}) that 
\begin{equation}                                                                         \label{nmcoord2}
\hat X = 
\sqrt{\hbar} \,  
                \left( {\mathsf S}_{\rm e}^\top {\mathsf C}_{[2]}^\dag \oplus 
                       {\mathsf S}_{\rm N}^\top {\mathsf C}_{[N]}^\dag \right) \hat Z,  
\end{equation} 
with $\hat Z = (\hat{\sf z}^\dag, \hat{z}^\dag)^\dag$. 
Finally, using the transformation (\ref{nmcoord2}) 
in the free Hamiltonian (\ref{hamfree}), one finds
\begin{equation}                                                                         \label{hamfree2}
\hat H_0 =  \frac{\hbar}{2} \, \hat Z^{\dagger} \! 
                              ( \Lambda_{\rm e} \oplus \Lambda_{\rm N} ) \hat Z 
         =  \frac{\hbar\Omega}{2} \, \hat {\sf z}^{\dagger}\hat{\sf z} +
            \frac{\hbar}{2} \, \hat {    z}^{\dagger} \! \Lambda_{\rm N} \hat{    z},                                
\end{equation}
which is the free Hamiltonian (\ref{hamfree}) written in the 
cre\-a\-ti\-on/annihilation representation of the normal mode coordinates.
%
%%%%%%%%%%%%%%%%%%%%%%%%%%%%%%%%%%%%%%%%%%%%%%%%%%%%%%%%%%%%%%%%%%%%%%%%%%%%%%%%%%%%%%%%%
\subsection{ Interaction Picture } 
%%%%%%%%%%%%%%%%%%%%%%%%%%%%%%%%%%%%%%%%%%%%%%%%%%%%%%%%%%%%%%%%%%%%%%%%%%%%%%%%%%%%%%%%%
Let us now move the dynamics to the interaction picture with respect to the free
Hamiltonian as given in Eq.~(\ref{hamfree2}). 
The LME in this picture acquires the following form
\begin{equation}                                                                         \label{lindblad2}
\frac{d \tilde \rho }{d t}  = 
         \frac{i}{\hbar}  [ \tilde \rho , \tilde H ] + \tilde{\mathcal{L}}(\tilde\rho),
\end{equation} 
with 
$\tilde H =    {\rm e}^{ \frac{i}{\hbar}\hat H_0 t} \, \hat H \, 
               {\rm e}^{-\frac{i}{\hbar}\hat H_0 t}  - \hat H_0$ and 
$\tilde \rho = {\rm e}^{ \frac{i}{\hbar}\hat H_0 t} \, \hat \rho \, 
               {\rm e}^{-\frac{i}{\hbar}\hat H_0 t} $. 
Also, all operators contained in $\mathcal L$ transform in the same way as $\hat \rho$, 
leading then to  $\tilde{\mathcal{L}}(\tilde\rho)$. By now, let us turn our attention 
to the Hamiltonian part. 

When moving to the interaction picture, the position operators of the oscillators in 
the chain and of the external ones, 
see (\ref{nmcoord2}), transform respectively according to 
\begin{equation}
\begin{aligned}
\tilde{q}_k & =  \tilde x_k = 
\sqrt{\hbar} \left( 
                    \mathsf S_{\rm N}^{\top} \mathsf{C}_{[N]}^{\dag} \tilde{z} 
             \right)_{\! k},   \,\,\,     1 \ge k \ge N ;                        \\
\tilde {\sf q}_{k} & =  \tilde {\sf x}_{k} =
\sqrt{\frac{\hbar}{2}} ( \tilde{\sf a}_k  + \tilde{\sf a}_k^{\dag} ), \,\,\, k = a,b.                                                           
\end{aligned}
\end{equation}
For what comes next, it is worth noticing that   
\begin{equation}
\tilde{q}_k^{2}  = \tilde x_k^2 = (\tilde x \tilde x^\dag)_{kk} = 
                    \hbar \left( 
                          \mathsf S_{\rm N}^{\top} \mathsf{C}_{[N]}^{\dag}
                          \tilde{z}\tilde{z}^{\dag} 
                          \mathsf S_{\rm N}^{\top} \mathsf{C}_{[N]}^{\dag}  
                          \right)_{kk}.                   
\end{equation}
Now, with all these in hand, we move $\hat H = \hat H_0 + \hat H_I$ 
to the interaction picture. 
By using  (\ref{hamint}) and (\ref{hamfree2}), one can see that the interaction 
picture Hamiltonian reads
\begin{eqnarray}                                                                         \label{hamint2} 
\tilde H &=& \frac{\epsilon}{4} (\tilde{q}_\alpha - \tilde{q}_a)^{2} + 
             \frac{\epsilon}{4} (\tilde{q}_\beta  - \tilde{q}_b)^{2}   \nonumber        \\
         &=& \frac{\hbar\epsilon}{4} 
             \left[    
                  \left( \mathsf S^{\!\top}\mathsf{C}^{\dag} 
                         \tilde{z}\tilde{z}^{\dag} 
                          \mathsf C \mathsf S                 \right)_{\alpha\alpha} +
                   \left( \mathsf S^{\!\top}\mathsf{C}^{\dag} 
                          \tilde{z}\tilde{z}^{\dag} 
                          \mathsf C \mathsf S                 \right)_{\beta\beta}     
            \right]                  \nonumber \\
&-& 
\frac{\hbar\epsilon}{2\sqrt{2}} 
\left[  
       \left(   \mathsf S^{\!\top}\mathsf{C}^{\dag} \tilde{z}\right)_\alpha  
            (   \tilde{\sf a}_a  + \tilde{\sf a}_a^{\dag} ) +
       \left(   \mathsf S^{\!\top}\mathsf{C}^{\dag} \tilde{z}\right)_\beta  
            ( \tilde{\sf a}_b  + \tilde{\sf a}_b^{\dag} ) 
\right] \nonumber \\
&+& \frac{\hbar\epsilon}{8} ( \tilde{\sf a}_a  + \tilde{\sf a}_a^{\dag} )^{2} +  
   \frac{\hbar\epsilon}{8} ( \tilde{\sf a}_b  + \tilde{\sf a}_b^{\dag} )^{2},  
\end{eqnarray}
where we dropped the indexes of $\mathsf S_{\rm N} $ and of $\mathsf{C}_{[N]}$ 
for notation simplicity, and
\begin{equation}
\begin{aligned}                                                                         \label{zint}                  
\tilde{z}_k &:= {\rm e}^{\frac{i}{\hbar}\hat H_0 t} \, 
                                         \hat z_k \, 
                                         {\rm e}^{-\frac{i}{\hbar}\hat H_0 t}  
                                       = {\rm e}^{i \phi_k t} \hat z_k  
\,\,\,  (k = 1,...,2N),         \\                               
\tilde{\sf a}_k &:= {\rm e}^{-i \Omega t} \hat{\sf a}_k 
\,\,\, (k = a,b) ,                                                                      
\end{aligned}
\end{equation}
with
\begin{equation}\label{nl}
\phi_k := \left\{ 
                 \begin{array}{r}
                 -\varsigma_k,  \,\,\,     \text{if} \,\,\,  k \le N \\
                  \varsigma_{k-N}, \,\,\,  \text{if} \,\,\,  k > N  
                 \end{array}
         \right. .                                 
\end{equation}
Notice that $\tilde z_k$ is  
$\tilde{a}_k = \hat{a}_k {\rm e}^{-i\varsigma_k t}$ provided $k \le N $ 
or its hermitian conjugate otherwise.
%

%%%%%%%%%%%%%%%%%%%%%%%%%%%%%%%%%%%%%%%%%%%%%%%%%%%%%%%%%%%%%%%%%%%%%%%%%%%%%%%%%%%%%%%%%
\subsection{ RWA and Effective Hamiltonian } \label{eh}
%%%%%%%%%%%%%%%%%%%%%%%%%%%%%%%%%%%%%%%%%%%%%%%%%%%%%%%%%%%%%%%%%%%%%%%%%%%%%%%%%%%%%%%%%
Under certain circumstances, fast oscillating terms in the interaction picture 
Hamiltonian are negligible and dropping these terms is what is called RWA.
In the Appendix~\ref{rwa}, we quantitatively describe such conditions for a 
prototype system of two coupled oscillators. This guides us in the application 
of the RWA for the present system of interest.  

Since the free Hamiltonian (\ref{hamfree2}) is the sum of $N+2$ 
non-interacting oscillators, each of its eigenvectors consists of tensor products of 
Fock states of each oscillator, {\it i.e.},  
\begin{equation}                                                                         \label{psi1}
| \Psi \rangle = |\mathfrak{n}_a,\mathfrak{n}_b, 
                  \mathfrak{n}_1,...,\mathfrak{n}_N  \rangle, 
\end{equation}
where $ \{ | \mathfrak{n}_k \rangle ; k=1,...,N \} $ 
are eigenstates of $ \hat{a}^{\dag}_k \hat{a}_k $ and 
$ \{ |\mathfrak{n}_k\rangle ; k=a,b \} $, the eigenstates of 
$ \hat{\sf a}^{\dag}_k \hat{\sf a}_k $. 
All transitions between eigenstates of the free Hamiltonian 
will be promoted by $\tilde{H_I}$, {\it i.e.}, by (\ref{hamint2}). 
Allowed transitions $| \Psi' \rangle\leftrightarrow|\Psi \rangle$, 
in the scope of first order perturbation theory, are those with
$\langle \Psi | \tilde H_I | \Psi' \rangle \ne 0 $.  
First-order time-dependent perturbation theory \cite{cohen} predicts that 
resonant or quasi-resonant transitions or, equivalently, those driven by 
static or slowly varying time dependent terms in the interaction picture Hamiltonian,
take place with higher probability when compared to transitions driven by 
the rapidly oscillating terms. 
The RWA consists of discarding the highly oscillating terms in the interaction 
Hamiltonian that are responsible for negligible transition amplitudes when compared 
to other terms that are static or oscillate slowly in time. Appendix~\ref{rwa} 
is dedicated to considerations about this approximation. Since the reasoning 
for usefulness of the RWA is that of the first-order perturbation theory, 
we must assure that $H_I$ is weak compared to  $H_0$. 
In our problem, this means that the interaction of the external oscillators 
with the network is weak, and this is guaranteed provided  
$\epsilon \ll \Omega,\varsigma_k$ ($k = 1,...,N$).  
 
Let us start with the case where the spectrum (\ref{nettw1}) is {\it non-degenerate}. 
Mathematically, that corresponds to $\varsigma_j = \varsigma_k$ if and only if $j = k$. 
In the system considered here, a  interesting scenario appears when the frequency of 
the external oscillators  $\Omega$ is close to the frequency of one of 
the normal modes of the network, let us say the $m^{\text{th}}$ mode $(1 \le m \le N)$, 
{\it i.e.}, $\Omega=\varsigma_m$.
In this case, if (\ref{zint}) is substituted in (\ref{hamint2}), 
the RWA can be performed using the recipe 
\begin{equation}                                                                         \label{rwaap} 
\begin{aligned}
&\tilde{z}_k\tilde{z}_l^{\dag} = 
\hat{z}_k\hat{z}_l^{\dag} \, {\rm e}^{i(\phi_k - \phi_l)t} 
\xrightarrow{ {\rm RWA} } \hat{z}_k\hat{z}_l^{\dag} \, \delta_{kl},          \\
&\tilde{\sf a}^{2}_l = \hat{\sf a}^{2}_l {\rm e}^{- 2 i  \Omega t}  
\xrightarrow{ {\rm RWA} } 0,\,\,\,
\tilde{\sf a}^{\dag 2}_l = \hat{\sf a}^{\dag 2}_l {\rm e}^{2 i \Omega t} 
\xrightarrow{ {\rm RWA} } 0 ,                                                \\  
&\tilde{z}_k \tilde{\sf a}^{\dag}_l = 
\hat{z}_k\hat{\sf a}_l^{\dag} \, {\rm e}^{i(\phi_k + \Omega)t} 
\xrightarrow{ {\rm RWA} } \hat{z}_k\hat{\sf a}_l^{\dag} \, 
\delta_{km},                                                                 \\
&\tilde{z}_k \tilde{\sf a}_l = 
\hat{z}_k\hat{\sf a}_l \, {\rm e}^{i(\phi_k - \Omega )t} 
\xrightarrow{ {\rm RWA} } \hat{z}_k\hat{\sf a}_l \, \delta_{k\, m+N} .   
\end{aligned}
\end{equation}
One can see that, under resonance condition $\Omega=\varsigma_m$ and weak 
interaction of external oscillators with the network, 
both necessary conditions for RWA to work well,  oscillators $a$ and $b$ 
couple essentially only with the $m^{\text{th}}$ normal mode.
Since modes other than $m$ follow free evolution, 
we do not include them in the effective description of the dynamics. 
Taking all these in to account, we can arrive at the following effective 
Hamiltonian\footnote{%%%%%%%%%%%%%%%%%%%%%%%%%%%%%%%%%%%%%%%%%%%%%%%%%%%%%%%%%%
It can be useful to write the elements of $\mathsf C_{[n]}$ 
in (\ref{ctrans}) as \\
$(\delta_{jk} + i \delta_{j k+n} + i \delta_{j k-n} )/\sqrt{2}$. 
%%%%%%%%%%%%%%%%%%%%%%%%%%%%%%%%%%%%%%%%%%%%%%%%%%%%%%%%%%%%%%%%%%%%%%%%%%%%%%%
}  
\begin{eqnarray}                                                                         \label{hamint3a} 
\tilde H_{\rm eff}^{(m)}  & =& \frac{\hbar\epsilon}{4}\mathcal{C}_{m}^{\alpha \beta} 
                                \hat a_m^\dagger\hat a_m                                     
+ \frac{\hbar\epsilon}{4} 
( \hat{\sf a}_a^{\dag}\hat{\sf a}_a + \hat{\sf a}_b^{\dag}\hat{\sf a}_b )          \\
&-& \frac{\hbar\epsilon}{4} 
\left[ \mathcal{D}_{\!m}^{\alpha}         \hat a_m \hat{\sf a}_a^{\dag} 
+      \bar{\mathcal{D}}_{\!m}^{\alpha}   \hat a_m^{\dag} \hat{\sf a}_a 
+      \mathcal{D}_{\!m}^{\beta}          \hat a_m\hat{\sf a}_b^{\dag} 
+      \bar{\mathcal{D}}_{\!m}^{\beta}    \hat a_m^{\dag} \hat{\sf a}_b \right],   \nonumber                                  
\end{eqnarray}
with
\begin{equation}                                                                         \label{coefint}
\begin{aligned}
&\!\!\!\!\! %
\mathcal{C}_{m}^{\alpha \beta} := 
                        ( \mathsf S_{ m \alpha }^2 + \mathsf S_{m+N \alpha}^2 + 
                          \mathsf S_{ m \beta  }^2 + \mathsf S_{m+N \beta }^2),          \\
&\!\!\!\!\! %
\mathcal{D}_{m}^{\mu} := ( \mathsf S_{m \mu } - i \mathsf S_{ m+N \, \mu } ), \,\,\, 
\bar{\mathcal{D}}_{m}^{\mu} := ( \mathsf S_{m \mu } + i \mathsf S_{ m+N \, \mu } ).  
\end{aligned}
\end{equation}

If the symplectic spectrum possess some degree of degeneracy, 
let us say $\varsigma_m = \varsigma_n$ for some $1\leq n,m\leq N$, 
and we tune $\Omega = \varsigma_m = \varsigma_n$, 
then scheme (\ref{rwaap}) is no longer valid. It must be modified to
\begin{equation}                                                                         \label{rwaap3} 
\begin{aligned}
&\tilde{z}_k\tilde{z}_l^{\dag} 
\xrightarrow{ {\rm RWA} } \hat{z}_k\hat{z}_l^{\dag} \,  
( \delta_{kl} + \delta_{km}\delta_{ln} + \delta_{kn}\delta_{lm} + \\
& \,\,\,\,\,\,\,\,\,\,\,\,\,\,\,\,\,\,\,\,\,\,
  \,\,\,\,\,\,\,\,\,\,\,\,\,\,\,\,\,\,\,\,\,
\delta_{k\,m+N}\,\delta_{l\,n+N} + \delta_{k\,n+N}\,\delta_{l\,m+N}),     \\
&\tilde{\sf a}^{2}_l = \hat{\sf a}^{2}_l {\rm e}^{- 2 i  \Omega t}   
\xrightarrow{ {\rm RWA} } 0,\,\,\,
\tilde{\sf a}^{\dag 2}_l = \hat{\sf a}^{\dag 2}_l  {\rm e}^{2 i \Omega t} 
\xrightarrow{ {\rm RWA} } 0 ,  \\  
&\tilde{z}_k \tilde{\sf a}^{\dag}_l
\xrightarrow{ {\rm RWA} } \hat{z}_k\hat{\sf a}_l^{\dag} \,
( \delta_{km} + \delta_{kn} ),                                         \\
&\tilde{z}_k \tilde{\sf a}_l 
\xrightarrow{ {\rm RWA} } \hat{z}_k\hat{\sf a}_l \,     
(\delta_{k\,m+N}+\delta_{k\, n+N}). 
\end{aligned}
\end{equation}
The additional terms, in comparison with (\ref{rwaap}), 
bring new elements for the dynamics the system.  Now, the situation involves 
the free dynamics of the degenerate modes, 
their mutual coupling, and their coupling with the external oscillators.
Following the same steps as before, we can now write an effective Hamiltonian 
for the system as
\begin{equation}
\begin{aligned}                                                                         \label{hameff2} 
&\tilde H_{\rm eff}^{(m,n)}  = \tilde H_{\rm eff}^{(m)} + \tilde H_{\rm eff}^{(n)} 
- \frac{\hbar\epsilon}{4} ( \hat{\sf a}_a^{\dag}\hat{\sf a}_a + 
        \hat{\sf a}_b^{\dag}\hat{\sf a}_b )    \\
&+\frac{\hbar\epsilon}{4} 
     ( \mathsf S_{m \alpha}\mathsf S_{n \alpha} + 
       \mathsf S_{m+N \alpha}\mathsf S_{n+N \alpha} ) 
     (\hat a_m^\dagger\hat a_n + \hat a_n^\dagger\hat a_m) \\
&+\frac{\hbar\epsilon}{4}( \mathsf S_{m \beta}\mathsf S_{n \beta} + 
       \mathsf S_{m+N \beta}\mathsf S_{n+N \beta}  ) 
     (\hat a_m^\dagger\hat a_n + \hat a_n^\dagger\hat a_m) \\
&+i\frac{\hbar\epsilon}{4} 
     ( \mathsf S_{m \alpha}\mathsf S_{n+N \alpha} - 
       \mathsf S_{m+N \alpha}\mathsf S_{n \alpha})
     (\hat a_m^\dagger\hat a_n - \hat a_n^\dagger\hat a_m) \\ 
&+i\frac{\hbar\epsilon}{4} 
     (\mathsf S_{m \beta}\mathsf S_{n+N \beta} - 
       \mathsf S_{m+N \beta}\mathsf S_{n \beta}  ) 
     (\hat a_m^\dagger\hat a_n - \hat a_n^\dagger\hat a_m), 
\end{aligned}
\end{equation}
where $\tilde H_{\rm eff}^{(k)}$ is given in (\ref{hamint3a}) for $k=m,n$. 

One may notice that, if $\mathsf S_{ m \alpha } = \mathsf S_{ m + N \alpha } = 0 $ 
in (\ref{hamint3a}) or in (\ref{hameff2}), oscillator $a$ will be decoupled from the 
$m^\text{th}$ normal mode of the network. Physically, 
position $\alpha$ would correspond to a {\it node} 
of the normal mode. 
In this situation, the coupling of the external oscillator $a$ with the normal mode $m$ 
takes place only for higher orders in $\epsilon$. Naturally, there is an analogous 
conditions for oscillator $b$.

For completeness, we would also like to mention the possibility of the external 
oscillators interacting with more than one member of the network. 
Suppose that oscillator $b$ is coupled to both oscillators $\beta$ and $\beta'$ 
simultaneously, {\it i.e.}, the interaction Hamiltonian is now 
\begin{equation}                                                                         \label{hamint4} 
\hat H'_I = \hat H_I + 
            \frac{\epsilon'}{4} \left(\hat q_{\beta'} - \hat {\sf q}_b \right)^2   
\end{equation}
with $\hat H_I$ given in (\ref{hamint}). 
All the calculations follow as before provided one now imposes $\epsilon' \ll \Omega$, 
in order to fulfill the RWA requirements. 
Specially, we would like to emphasize that $\mathsf S_{0}$ defined above Eq.~(\ref{hamtw}) 
remains the same. After the calculations, the result is
\begin{equation}                                                                         \label{hamint4a} 
\begin{aligned}
\tilde H_{\rm eff}^{'(m)}   = & \, \tilde H_{\rm eff}^{(m)} + 
\frac{\hbar\epsilon'}{4}\mathcal{E}_{m}^{\beta'} \hat a_m^\dagger\hat a_m                                     
+ \frac{\hbar\epsilon'}{4} \hat{\sf a}_b^{\dag}\hat{\sf a}_b           \\
& - \frac{\hbar\epsilon'}{4} 
\left[ 
       \mathcal{D}_{\!m}^{\beta'}          \hat a_m\hat{\sf a}_b^{\dag} 
+      \bar{\mathcal{D}}_{\!m}^{\beta'}    \hat a_m^{\dag} \hat{\sf a}_b \right],                                     
\end{aligned}
\end{equation}
with 
$\tilde H_{\rm eff}^{(m)}$ in (\ref{hamint3a}), 
$\mathcal{D}_{\!m}^{\beta'}$ in (\ref{coefint}), and 
$
\mathcal{E}_{m}^{\beta'} := \mathsf S_{ m \beta' }^2 + \mathsf S_{m+N \beta' }^2. 
$
Note also that if one wants to consider in (\ref{hamint}) the possibility 
of distinct couplings namely $\epsilon$ for $(a,\alpha)$ and $\epsilon'$ for $(b,\beta)$, 
the use of (\ref{hamint4a}) for the part referring to $(b,\beta)$ is the way to proceed.%

%%%%%%%%%%%%%%%%%%%%%%%%%%%%%%%%%%%%%%%%%%%%%%%%%%%%%%%%%%%%%%%%%%%%%%%%%%%%%%%%%%%%%%%%%
\subsection{ Thermal Baths and Effective Dynamics } \label{t}
%%%%%%%%%%%%%%%%%%%%%%%%%%%%%%%%%%%%%%%%%%%%%%%%%%%%%%%%%%%%%%%%%%%%%%%%%%%%%%%%%%%%%%%%%
The unavoidable incapacity of a perfect system isolation leads to the progressive 
destruction of quantum coherence. This kind of dynamics is commonly modeled by including 
appropriate non-unitary components in the equation of motion. 
We will specialize here in the case of local and independent thermal 
baths for each member of the system depicted in Fig.\ref{fig1system1}. 
This physical scenario corresponds to the use of (\ref{liouv}) with the choice  
\begin{eqnarray}                                                                         \label{lopthermal}
\hat L_{(k)} &=&  \sqrt{\hbar\zeta ({\bar n}_{\text{th}} + 1)} \, \hat A_k ,\nonumber\\
\hat L'_{(k)} &=& \sqrt{\hbar\zeta {\bar n}_{\text{th}}} \, \hat A_{k}^\dagger, 
\,\,\, (k = a,b,1,2, ...,N),
\end{eqnarray}
where $\hat A_k$ is the annihilation operator associated to the $k^{\text{th}}$ 
component of $\hat X$ , 
$\zeta \ge 0$ is the effective bath-oscillator coupling constant or relaxation 
rate of the system, and  
${\bar n}_{\text{th}}$ is a thermal occupation number, both taken here to be the same 
for all reservoirs.   
Note that, for the same $k$, two Lindblad operators must be simultaneously included 
(primed and unprimed).  
The prescribed Lindblad operators do not couple different oscillators, 
and this brings the matrix (\ref{decmatdef}) to a block structure 
\begin{equation}                                                                         \label{decmat} 
{\bf \Upsilon} =  {\bf \Upsilon}_{4} \oplus {\bf \Upsilon}_{2N},
\end{equation} 
with 
\begin{equation}                                                                         \label{decmat2}
{\bf \Upsilon}_{\! 2n} := 
\zeta({\bar n}_{\text{th}} + \tfrac{1}{2}) \mathsf I_{n} - 
\frac{i}{2} \zeta \mathsf J^{[n]}, 
\end{equation}
where $n$ equals $2$ or $N$. 

Now, the same transformation (\ref{sdt}), used to diagonalize $\hat H_0$ 
in (\ref{hamtw}), must be applied to the Lindblad operators defined in (\ref{lindef}). 
Consequently,
\begin{equation}                                                                         \label{lindint}
\hat {L}_{(k)} =  
\lambda_{(k)}^{\!\top} \mathsf J  \mathsf S_0^{\!\top}\hat Y
= [\mathsf S_0^{-\top} \lambda_{(k)} ]^{\!\top} 
      \mathsf J \hat Y.
\end{equation}
To see that, one must apply the sympleticity of $ \mathsf S_0$, {\it i.e.}, 
$\mathsf S_0^{\!\top} \! \mathsf J \mathsf S_0 = \mathsf J$. 
Notice that we used a compact notation $\mathsf S_0^{-\top}$ for 
$(\mathsf S_0^\top)^{-1}$.  
Given (\ref{decmatdef}), we see that
\begin{equation}                                                                         \label{dectransf} 
{\bf \Upsilon} \longrightarrow \mathsf S_0^{-\top} {\bf \Upsilon} \mathsf S_0^{-1},
\end{equation}
in such a way that (\ref{decmat}) becomes
\begin{eqnarray}                                                                         \label{decmat4}
\!\!\!\!\!\!\!\!\!\!\mathsf S_0^{-\top} {\bf \Upsilon} \mathsf S_0^{-1} = 
\zeta({\bar n}_{\text{th}} + \tfrac{1}{2}) 
(\mathsf S_0\mathsf S_0^{\top})^{-1}  
- \tfrac{i}{2} \zeta \left(\mathsf J^{[2]} \oplus \mathsf J^{[N]}\right), 
\end{eqnarray}
where we used the symplecticity of $\mathsf S_0$ again. 

Our aim is to provide the simplest description of the dynamics of oscillators $a$ and $b$ 
mediated by the network. At the Hamiltonian level, 
we already managed to do that when we arrived at an effective interaction involving just 
these oscillators and a few resonant normal modes.
For the Lindblad operators and covariance matrices, the description in terms of normal 
modes is reflected in (\ref{decmat4}). In general, the normal modes turn out to be 
interacting in spite of the fact that, looking at the individual oscillators, 
they interact with independent baths. 
One can say that the action of the baths, in the level of normal modes 
(which are collective operators), is non-local. The interaction of the normal modes 
in the non-unitary part of the dynamics comes 
from the mutual interactions of the individual oscillators in the Hamiltonian, {\it i.e.}, 
in the unitary part of the dynamics. Consequently,  (\ref{decmat4}) might not be as 
simple as (\ref{decmat2}), remembering that the latter refers to a description of local 
independent baths. 
From this, we see that the problem of obtaining a simple description of the open system 
dynamics is much more involved than the same problem in the closed unitary dynamics. 
So, together with the treatment of a general network, the possible simplifications for 
the open system dynamics we do next make our contribution of interest giving that 
previous studies treated only closed systems in a fixed 
simple topology \cite{plenio2005}. 

It is possible to envision some structural conditions that make (\ref{decmat4}) simpler.  
In other words, conditions that lead to local and independent baths for the collective 
normal modes. This basically concerns the form of the matrix 
$\mathsf S_0\mathsf S_0^{\top}$ (topology), or the form of $\lambda_{(k)}$ 
(system-bath interaction). Let us start with the first condition which will be 
illustrated with an example in Sec.~\ref{elc}.

Direct inspection of  (\ref{decmat4}) reveals that the baths for the normal modes 
will naturally be local and independent provided 
$\mathsf S_0\mathsf S_0^{\top}$ becomes a diagonal matrix. One possibility is 
$\mathsf S_0\mathsf S_0^{\top}=\mathsf I_{4}\oplus\mathsf I_{2N}$ what would lead 
precisely to a form like (\ref{decmat2}), corresponding to interaction with local 
independent baths. When $\mathsf S_0\mathsf S_0^{\top}$ is diagonal but not the identity 
matrix, each mode will still see a local reservoir but it will not necessarily be thermal. 
In this case there is a weighted mix of creation and annihilation operators 
characteristic of a squeezed reservoir.
Given the quadratic Hamiltonian in (\ref{hamfree}), the results in Sec.\ref{sf} will be helpful 
to determine if the action of the reservoirs will be decoupled or not. 
Provided the blocks of $\bf H_{\rm N}$ satisfy the conditions in (\ref{condtheo}), 
it will be suitable for symplectic diagonalization by a matrix $\mathsf S_{\rm N}$ such that 
$\mathsf S_{\rm N} \mathsf S_{\rm N}^\top$ is a diagonal matrix. 
Considering that the same is true for the blocks of $\bf H_{\rm e}$ in (\ref{hamfree}), 
then $\mathsf S_0$ in (\ref{hamtw}) will be such that 
$\mathsf S_{0} \mathsf S_{0}^\top$ is diagonal.

For resonance of $a$ and $b$ with a non-degenerate normal mode $m$, we define
\begin{equation}                                                                         \label{redvec}
\check x = ( \hat{\sf q}_a, \hat{\sf q}_b, \hat y_m,                      
	     \hat{\sf p}_a, \hat{\sf p}_b,\hat y_{m+N})^{\dag},            
\end{equation}
from which we may write the effective Hamiltonian (\ref{hamint3a}) as 
\begin{equation}                                                                         \label{hesseff}             
\hat H_{\rm eff}^{(m)} = 
\frac{1}{2} \check x^\dag {\bf H}_{\rm eff} \check x = 
\frac{\epsilon}{8} \check x^\dag 
\begin{pmatrix}
{\bf H_q} & {\bf C_{\bf qp}} \\
{\bf C^{\!\top}_{\bf qp}} & {\bf H_p}
\end{pmatrix}
\check x 
\end{equation}
with 
\begin{equation}                                                                         \label{hesseffa}
{\bf H_q} = {\bf H_p} = 
\left[ \begin{array}{ccc}
1 & 0 &  - \mathsf S_{m \alpha }  \\
0 & 1 &  - \mathsf S_{m \beta }  \\
- \mathsf S_{m \alpha } & - \mathsf S_{m \beta } &  
\substack{    ( \mathsf S_{m \alpha}^{2}+\mathsf S_{m \beta}^{2} +  \\ 
                \mathsf S_{m+N \, \alpha }^{2} + \mathsf S_{m+N \, \beta }^{2} )   } \\
 \end{array} \right],
\end{equation}
and 
\begin{equation}                                                                         \label{hesseffb}
{\bf C_{\bf qp}} = 
\left[ \begin{array}{ccc}
0 &  0 &  - \mathsf S_{m+N \, \alpha }  \\
0 &  0 &  - \mathsf S_{m+N \, \beta }   \\
\mathsf S_{m+N \, \alpha } & \mathsf S_{m+N \, \beta } & 0  \\
 \end{array} \right] .
\end{equation}
Using (\ref{cmdef}), a CM based on $\check x$ can be built and, just like (\ref{cmev}), 
it will evolve according to
\begin{equation}                                                                         \label{cmeveff}
\frac{d}{d t} \check{ \mathbf V } = 
\check{ \bf \Gamma }\check{ \mathbf V } + \check{ \mathbf V } \check {\bf \Gamma}^\top 
+ \check{\bf D} 
\end{equation}
with  
\begin{equation}                                                                         \label{decmateff} 
\check{\bf \Gamma } := 
\mathsf J^{[6]} {\mathbf H}_{\rm eff} 
 - \frac{\zeta}{2} \mathsf I_{6} , \,\,\,                                                                           
\check{\bf D}  :=  \hbar \zeta({\bar n}_{\text{th}} + \tfrac{1}{2}){\mathsf D}, 
\end{equation}
where $\mathsf D$ will be a $6 \times 6$ diagonal matrix 
since $\mathsf S_0\mathsf S_0^{\top}$ is considered to be diagonal.
Solution  (\ref{cmsol}) applied to this effective description reads
\begin{equation}                                                                         \label{cmsol1}
\check{\bf V}(t)  =  {\rm e}^{\check{\bf \Gamma} t} \, \check{\bf V}\!_{0}  \, 
            { \rm e}^{ \check{\bf \Gamma}^{\! \top} t } +
 \int_0^t \! dt^\prime \, 
               {\rm e}^{\check{\bf \Gamma} t^\prime} \, 
                  \check{\bf D}  \,
               {\rm e}^{\check{\bf \Gamma}^{\! \top} t^\prime}  \, ,
\end{equation}
with  ${\exp[\check{\bf \Gamma}t] = {\rm e}^{-\zeta t/2}} \mathsf E (t)$, where
\begin{equation}                                                                         \label{rwssimp}
\mathsf E (t) = \exp\left[ \mathsf J
                                      {\mathbf H}_{\rm eff}\, t      \right] 
                                     \in {\rm Sp}(6,\mathbb R).
\end{equation}
This represents a huge simplification to the original problem which is to describe 
the open system dynamics of $a$ and $b$ when they interact 
with a network of $N$ oscillators. This is especially true for big networks.
 
For a {\it degenerate} mode frequency, it suffices to build a  vector just like 
(\ref{redvec}) but now containing
all the degenerate modes. From this, one can proceed as in the non-degenerate case. 
For example, if the symplectic eigenvalue is two-fold degenerate, say modes $m$ and $n$, 
we define 
\begin{equation}                                                                         \label{degvec}
\check x = 
(\hat{\sf q}_a, \hat {\sf q}_b, \hat y_m,\hat y_n,
\hat {\sf p}_a, \hat {\sf p}_b, \hat y_{m+N},\hat y_{n+N})^{\dag},            
\end{equation}
and  ${\bf H}_{\rm eff}$, $\check{\bf \Gamma }$ and 
$\check{\bf D}$ will be $8 \times 8$ matrices. An example will be given in Sec.\ref{d}.

Let us now present a second condition allowing the description of the normal modes as 
subjected to local and non-interacting baths. This will happen whenever $ \lambda_{(k)}$ 
appearing in (\ref{lindef}) is of the special form 
$\lambda_{(k)} = \mathsf S_0^\top \mu_{(k)}$, where $\mathsf S_0$ 
is the symplectic matrix diagonalizing the Hamiltonian (\ref{hamtw}),
and $\mu_{(k)}$ corresponds to local thermal baths. In other words, $\mu_k$  
is determined from $\hat L_k =  \mu_{(k)}^{\!\top} \mathsf J \hat{X} $ 
with $\hat L_k$ given by (\ref{lopthermal}). 
Under these circumstances, 
the transformed matrix $\mathsf S_0^{-\top} {\bf \Upsilon} \mathsf S_0^{-1}$ 
in (\ref{dectransf}) assumes the form (\ref{decmat}) as a direct 
consequence of the symplectic property of $\mathsf S_0$:
\begin{equation}                                                                          \label{}
\hat L_k =  \lambda_{(k)}^{\!\top} \mathsf J \hat{X} = 
(\mathsf S_0^\top \mu_{(k)})^{\top} \mathsf J \mathsf S_0^\top \hat Y =
\mu_{(k)}^{\top} \mathsf J \hat Y.
\end{equation}
Since $\mu_{(k)}$ corresponds to local thermal baths, we achieved our goal. 

If we give up the requirement of having local baths, 
there is still other possibilities to attain an effective LME involving just a 
few degrees of freedom. For instance, when the structure of the network is such that 
the transformed $\Upsilon$ in (\ref{decmat4}) only interconnects the resonant  
oscillators, the effective dynamics will still only involve themselves, 
but possibly in a non-local way. 

%%%%%%%%%%%%%%%%%%%%%%%%%%%%%%%%%%%%%%%%%%%%%%%%%%%%%%%%%%%%%%%%%%%%%%%%%%%%%%%%%%%%%%%%%
\section{Example: Linear Chain}\label{elc}                        %%%%%%%%%%%%%%%%%%%%%%%
\begin{figure}[htpb!]                                                      
\includegraphics[width=8cm]{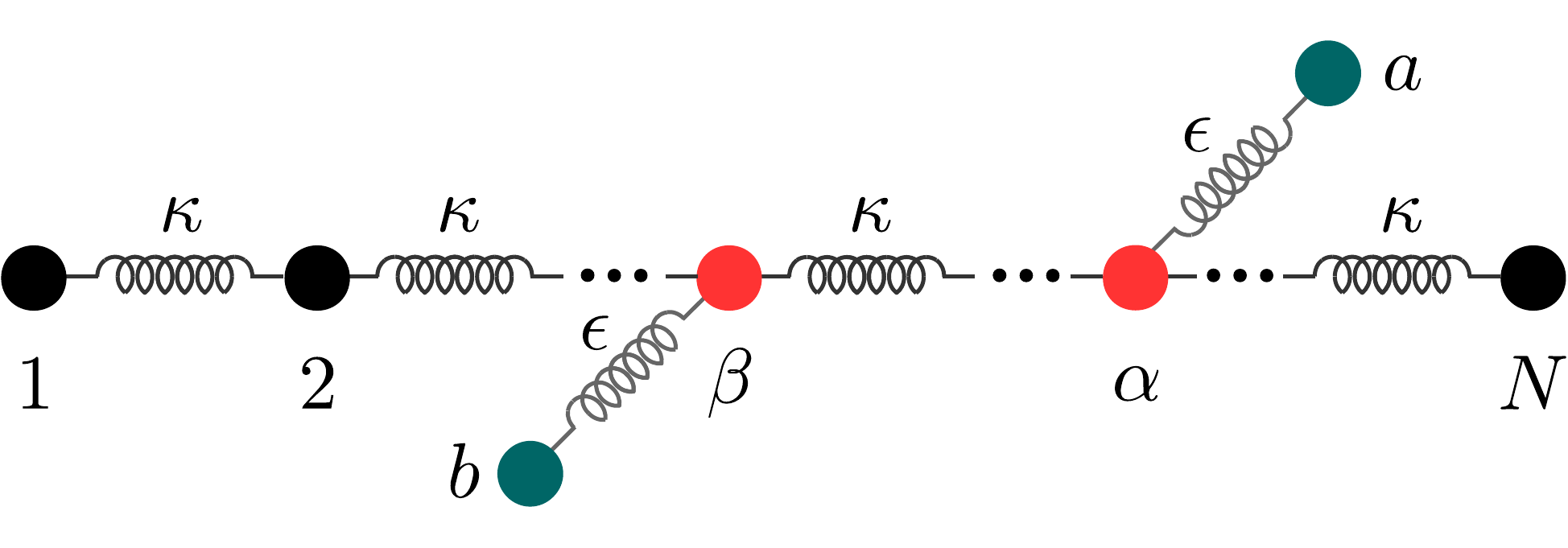} 
\caption{%
Chain of $N$-coupled harmonic oscillators as a network 
where oscillators $a$ and $b$ are attached at positions 
$\alpha$ and $\beta$, respectively. 
Coupling constants $\kappa$ and $\epsilon$ 
refer to Hooke-like forces.
}                                                                                        \label{fig2system2}                            
\end{figure}

%%%%%%%%%%%%%%%%%%%%%%%%%%%%%%%%%%%%%%%%%%%%%%%%%%%%%%%%%%%%%%%%%%%%%%%%%%%%%%%%%%%%%%%%%
Consider a chain of $N$ harmonic oscillators with frequency $\omega$ and 
coupled by springs (Hooke's law) with coupling constant $\kappa$, 
as depicted in Fig.\ref{fig2system2}. 
The external oscillators $a$ and $b$ couples respectively with the 
$\alpha^\text{th}$ and $\beta^\text{th}$ 
oscillators of the chain as in (\ref{hamint}) and have frequency $\Omega$. 

%%%%%%%%%%%%%%%%%%%%%%%%%%%%%%%%%%%%%%%%%%%%%%%%%%%%%%%%%%%%%%%%%%%%%
%%%%%%%%%%%%%%%%%%%%%%%%%%%%%%%%%%%%%%%%%%%%%%%%%%%%%%%%%%%%%%%%%%%%% 

The free Hamiltonian (\ref{hamfree}) for this particular configuration is defined with
\begin{equation}                                                                         \label{hesschain}                                      
{\mathbf H}_{\rm e} = \Omega \mathsf I_4, \,\,\, 
{\mathbf H}_{\rm N} = {\bf Q} \oplus {\omega \, \mathsf I_N }, 
\end{equation}
where ${\bf Q}$ is a $N \times N$ potential matrix whose elements are given by
\begin{equation}                                                                         \label{potchain}
{\bf Q}_{\! j k}  =  
    ( \omega + \kappa ) \delta_{jk} 
   - \tfrac{\kappa}{2} (\delta_{j1}\delta_{1 k} + \delta_{j n}\delta_{n k} +   
                        \delta_{j k\pm1} ).
\end{equation}
Notice that we are taking $M\Omega=1$ as discussed in Sec.\ref{di}.

The matrix ${\bf Q}$ in (\ref{potchain}) is tridiagonal and symmetric what implies that 
it can be diagonalized by an orthogonal transformation \cite{kulkarni}. In particular,
$ {\pmb O} {\bf Q} {\pmb O}^\top  =   {\rm Diag}(h_1,...,h_N )  $,
with
\begin{equation}                                                                         \label{toep}
\begin{aligned}
{\pmb O}_{\! j k} & =  
\sqrt{\frac{2-\delta_{j1}}{N} } \, \cos \tfrac{(j-1)(2k-1) \,\pi}{2 N},   \\
h_k & =  (\omega + \kappa)  -\kappa  \cos\tfrac{(k -1)  \,\pi}{ N}.   
\end{aligned}
\end{equation}

Now we proceed to reveal the normal modes of the chain. 
Following (\ref{tw2}), we calculate
\begin{equation}
{\rm Spec}(\mathsf J {\bf H}_{\rm N})  = (i{\varsigma}_1,...,i{\varsigma}_N,
                                   -i{\varsigma}_1,...,-i{\varsigma}_N),
\end{equation}
with
\begin{equation}                                                                         \label{sesmc}
{\varsigma}_k = \sqrt{ \omega (\omega + \kappa) - 
                       \omega \kappa \cos\tfrac{(k - 1)  \,\pi}{ N} } ,
\,\,\, 
k = 1,...,N . 
\end{equation}
The set (\ref{sesmc}) defines the symplectic spectrum (\ref{nettw1}) 
and is indeed not degenerate. 
The symplectic spectrum of the external oscillators is  
$\Lambda_{\rm e} = \Omega \mathsf I_4$.  
With these in hand, we are able to construct $\mathsf S_0$ which is the 
symplectic matrix (\ref{hamtw}) that diagonalizes  ${\bf H}_{\rm e}\oplus{\bf H}_{\rm N}$. 
By doing that,
the free Hamiltonian (\ref{hamfree}) with ${\bf H}_{\rm e}$ and ${\bf H}_{\rm N}$ will be 
indirectly diagonalized when written in terms of the normal modes. 
We start by conveniently writing $\mathsf S_0$ as  
\begin{eqnarray}                                                                         \label{sympdiag}
\mathsf S_0 =  \mathsf I_4 \oplus \mathsf S, 
\end{eqnarray}
being $\mathsf S$ the matrix that performs the simplectic diagonalization of 
${\bf H}_{\rm N}$. 
Considering 
${\bf M} = {\bf H}_{\rm N} = {\bf Q} \oplus {\omega \, \mathsf I_N }$ in (\ref{tw2a}), 
one can show that $O = {{\pmb O} \oplus {\pmb O} }$ 
is a solution of (\ref{must}) with $\pmb O$ defined in (\ref{toep}).  
By explicitly working with (\ref{tw2a}), 
it is now easy to show that
$
\mathsf S = {\pmb S} {\pmb O} \! \oplus \! {\pmb S}^{-1} {\pmb O}
$
with 
\begin{equation}
{\pmb S} = {\rm Diag}\left(\sqrt[4]{  \tfrac{\omega}{ h_1 } },...,
                           \sqrt[4]{\tfrac{\omega}{ h_N } }     \right),  
\end{equation}
for $h_k$ defined in (\ref{toep}).
Also, it may be useful to note that 
\begin{equation}                                                                         \label{aux}
{\mathsf S}_{m \mu} = 
\sqrt{ \frac{ \omega}{ \varsigma_m}} \,  \pmb{O}_{ \! m \mu}, \,\,\, 
{\mathsf S}_{m + N \mu} = 0 \,\,\,\,\, ( \mu = \alpha, \beta). 
\end{equation}%

When the external oscillators are put in resonance with the $m^{\text{th}}$ 
normal mode, {\it i.e.}, $\varsigma_m = \Omega$, 
the following effective Hamiltonian is obtained with application of (\ref{hamint3a})
and (\ref{aux})
\begin{equation}                                                                         \label{hameff} 
\begin{aligned}
&\tilde H_{\rm eff}^{(m)}  = 
\frac{\hbar\epsilon\omega}{4 { \varsigma_m }} 
\left(  \pmb{O}_{\! m \alpha}^{2}  +  \pmb{O}_{\! m \beta }^{2} \right) 
\hat a_m^\dagger\hat a_m  
+ \frac{\hbar\epsilon}{4} ( \hat{\sf a}^\dagger_a \hat{\sf a}_a +  
\hat{\sf a}^\dagger_b \hat{\sf a}_b   )  \\ 
& -\frac{\hbar\epsilon\sqrt{\omega}  }{4 \sqrt{ \varsigma_m} } 
\left[ 
\pmb{O}_{ \! m \alpha }( \hat a_m \hat{\sf a}_a^{\dag} + \hat a_m^{\dag} \hat{\sf a}_a)  
+ \pmb{O}_{ \! m \beta } ( \hat a_m \hat{\sf a}_b^{\dag} + \hat a_m^{\dag} \hat{\sf a}_b) 
\right].   
\end{aligned}
\end{equation}
Now a few important remarks. First, one can clearly see that the resonances are not 
equivalent. For example, if the resonant mode is chosen to be $m = 1$, 
that is $\Omega = \varsigma_1$, the dynamics will be independent of the positions 
that the external oscillators are connected to the chain (translational invariance). 
In other words, there is no dependency on  $\alpha$ and $\beta$ 
(see Fig.~{\ref{fig2system2}), and this follows from 
$\pmb{O}_{\! 1 \mu } = 1/\sqrt{N},\,\, \forall  \mu$, see Eq.~(\ref{toep}). 
For other resonances ( $\Omega = \varsigma_m$, $m \ne 1$), 
translational invariance is broken and the dynamics will drastically depend 
on $\alpha,\beta$. 
For instance, if $\alpha = k N/(2m -2) + 1/2 $ with $k \in \mathbb Z^\ast $,  
then $\pmb O_{m \alpha } = 0$, the external oscillator $a$ is effectively decoupled 
from the chain --- The aforementioned positions $\alpha$ are nodes (zero amplitude) 
of  high frequency modes ($m > 1$) and this results in decoupling. 
As a final remark, only when resonance is set with mode $m=1$, 
the closed chain considered in \cite{plenio2005} is equivalent to the open chain 
treated here, {\it i.e.}, both topologies can be described by Hamiltonian 
$\tilde H_{\rm eff}^{(1)}$   

Now we turn our attention to the interaction with the environment. 
One can see from (\ref{sympdiag})  that  
\begin{equation}                                                                         \label{restrans}                                                                       
\begin{aligned}
\mathsf S_0\mathsf S_0^{\top} &= 
\mathsf I_4 \oplus {\pmb S}^{2} \oplus {\pmb S}^{-2}   \\
& = \mathsf I_4 \oplus 
                      {\rm Diag}\!\left( \tfrac{\omega}{ \varsigma_1 },...,
                                         \tfrac{\omega}{ \varsigma_N }    ,
                                         \tfrac{\varsigma_1 }{ \omega },...,
                                         \tfrac{\varsigma_N }{ \omega } 
                                                                              \right ),   
\end{aligned}                 
\end{equation}
and this leads (\ref{decmat4}) to a special form whose physical interpretation is 
that each mode will see only a single local squeezed reservoir, 
as discussed in Sec.\ref{t}.
Finally, the physical situation is that of an effective dynamics comprising only the 
external oscillators and the $m^\text{th}$ 
normal mode of the chain, these three subjected to local baths. Since modes other than $m$ 
follow free dissipative evolutions (decoupled from $a$, $b$, and $m$), they do not have 
to be included in the description, provided our interest is in the external oscillators.
Coming back to the position/momentum representation (\ref{hesseff}), we obtain  
\begin{equation}                                                                         \label{hameffinal}
\tilde H_{\rm eff}^{(m)} = \tfrac{\epsilon}{8} \check x^\dag \mathbf{H_q}\oplus 
                                           \mathbf{H_q}\check x
\end{equation}
with 
\begin{equation}                                                                         \label{hesseff2}
\mathbf{H_q} = 
\left( \begin{array}{ccc}
1 & 0 & - \mathsf{S}_{ \! m \alpha }\\
0 & 1 &  
- \mathsf{S}_{ m \beta } \\
- \mathsf{S}_{ m \alpha } & 
- \mathsf{S}_{ m \beta } & 
(\mathsf{S}_{  m \alpha }^2 +  \mathsf{S}_{ m \beta }^2) 
 \end{array} \right),
\end{equation}
whose elements are given by (\ref{aux}). 
For (\ref{decmateff}), we have  
\begin{equation}                                                                         \label{decmateff2} 
\begin{aligned}                                                                
\check{\bf \Gamma } &:= 
\mathsf J^{[6]} \left( \mathbf{H_q}  \oplus \mathbf{H_q}  \right)
 - \frac{\zeta}{2} \mathsf I_{6} , \\                                                                           
 \check{\bf D} & :=  \hbar \zeta({\bar n}_{\text{th}}+ \tfrac{1}{2}) 
\left( \mathsf I_2 \oplus \tfrac{\varsigma_m}{\omega} \oplus 
       \mathsf I_2 \oplus \tfrac{\omega}{\varsigma_m}          \right),
\end{aligned}
\end{equation}
which, in association with the symplectic evolution (\ref{rwssimp}),  
\begin{equation}                                                                         \label{rwssimp2}
\mathsf E (t) = \exp\left[ \mathsf J( \mathbf{H_q}  \oplus 
                                      \mathbf{H_q} )\, t      \right] 
                                     \in {\rm Sp}(6,\mathbb R) \cap {\rm O}(6),
\end{equation}
allows one to obtain the time evolved CM according to the solution (\ref{cmsol1}). 
Detailed expressions for the matrix elements constituting $\mathsf E (t)$ 
can be found in Appendix~\ref{ma}.

%%%%%%%%%%%%%%%%%%%%%%%%%%%%%%%%%%%%%%%%%%%%%%%%%%%%%%%%%%%%%%%%%%%%%%%%%%%%%%%%%%%%%%%%%
%%%%%%%%%%%%%%%%%%%%%%%%%%%%%%%%%%%%%%%%%%%%%%%%%%%%%%%%%%%%%%%%%%%%%%%%%%%%%%%%%%%%%%%%%
\section{ Application: Energy Transport } \label{et}              %%%%%%%%%%%%%%%%%%%%%%%
%%%%%%%%%%%%%%%%%%%%%%%%%%%%%%%%%%%%%%%%%%%%%%%%%%%%%%%%%%%%%%%%%%%%%%%%%%%%%%%%%%%%%%%%%
%%%%%%%%%%%%%%%%%%%%%%%%%%%%%%%%%%%%%%%%%%%%%%%%%%%%%%%%%%%%%%%%%%%%%%%%%%%%%%%%%%%%%%%%%
The validity of the method developed so far is now carefully studied in a 
specific problem of importance for quantum technologies.  
This concerns the propagation of energy from a quantum system (oscillator $b$) 
to another (oscillator $a$) through a quantum bus (the network).  

%%%%%%%%%%%%%%%%%%%%%%%%%%%%%%%%%%%%%%%%%%%%%%%%%%%%%%%%%%%%%%%%%%%%%%%%%%%%%%%%%%%%%%%%%
\subsection{Non-degenerate normal modes}
%%%%%%%%%%%%%%%%%%%%%%%%%%%%%%%%%%%%%%%%%%%%%%%%%%%%%%%%%%%%%%%%%%%%%%%%%%%%%%%%%%%%%%%%%
Let us 
consider the propagation of energy between the oscillators $b$ and $a$ 
through the linear chain, as depicted in Fig.\ref{fig2system2}.  
For that, all oscillators are initially prepared in a tensor product of 
local vacuum states, 
except for $b$ which will be considered in a thermal state (TS). 
Thus, the CM (\ref{cmdef}) at $t=0$ for the global system reads  
\begin{equation}                                                                         \label{globalcm0}
{\bf V}_0 = 
{\bf V}_{\rm T} \oplus \frac{\hbar}{2} \mathsf I_{2N}, 
\end{equation}
in which the CM of the subsystem $(a,b)$ is given by  
\begin{equation}                                                                         \label{cm0}
{\bf V}_{\rm T } = \frac{\hbar}{2} 
\left( \begin{array}{cc}
       1 & 0  \\
       0 & 2 \bar{n}_b + 1       
       \end{array}  \right)  \oplus 
\frac{\hbar}{2} 
\left( \begin{array}{cc}
       1 & 0  \\
       0 & 2 \bar{n}_b + 1       
       \end{array}  \right), 
\end{equation}
where  
$\bar n_b \ge 0$ is the average number of thermal phonons initially 
in the oscillator $b$. Notice that oscillator $a$ is initially in the vacuum state. 
We will be interested in the dynamics of the average occupation number of $a$, 
and this can be extracted from the evolved CM as 
\begin{equation}                                                                         \label{mon}
{\bar n}_a := 
\langle \hat{\sf a}_a^\dag \hat{\sf a}_a \rangle_t  = 
%{\rm Tr}\left( \hat \rho(t) \hat A_a^\dag \hat A_a \right) = 
\tfrac{1}{2\hbar}\left[ {\bf V}_{\! 1 1 }(t) + {\bf V}_{\! 3 3 }(t) \right]
- \tfrac{1}{2}.
\end{equation}

Let us start with the ideal case of a perfectly isolated system. In this case, 
the evolution of ${\bf V}_0$ will be governed by (\ref{cmev}) 
with ${\bf \Upsilon}=0$ and ${\bf \Gamma } = \mathsf J \mathbf H$, {\it i.e.},
\begin{equation}                                                                         \label{cmsol2}
{\bf V}(t)  =  {\rm e}^{ \mathsf J {\bf H} t} \, {\bf V}\!_{0}  \, 
               {\rm e}^{- {\bf H}\mathsf J  t } .
\end{equation}
Note that ${\rm e}^{ \mathsf J {\bf H} t} \in {\rm Sp(2N+4,\mathbb R)}$. 
Despite the apparent simplicity of this formula, it involves exponentiation 
of ($2N+4)\times (2N+4)$ matrices, a difficult task depending on the magnitude of $N$. 
However, using the method developed in Sec.~\ref{ed}, one deals instead with 
exponentiation of $6\times 6$ matrices regardless of $N$:  
\begin{equation}                                                                         \label{cmsol3}
\check{\bf V}(t)  = {\mathsf E}(t) \, \check{\bf V}\!_{0}  \, {\mathsf E}^\top(t). 
\end{equation}
Of course, $N$ can not be considered arbitrarily big. 
In this case, the frequency of the modes will vary in a continuum, 
and this spoils the RWA~\cite{plenio2005}. 
Now, we evaluate the average occupation number of $a$ as
\begin{equation}                                                                         \label{moneff}
{\check n}_a := 
\tfrac{1}{2\hbar}\left[ \check{\bf V}_{\! 1 1 }(t) + \check{\bf V}_{\! 33 }(t) \right] 
- \tfrac{1}{2},
\end{equation}
which, after using the matrix elements presented in Appendix~\ref{ma}, become 
\begin{eqnarray}                                                                         \label{moneff2}
{\check n}_a = 
2 \bar{n}_b \, 
F({\mathsf S}_{m \alpha}^2 + {\mathsf S}_{m \beta}^2 + 1,\tfrac{\epsilon t}{4})  
\end{eqnarray}
with 
\begin{equation}                                                                         \label{funF}
F(\chi,\!\tau) \! = \!  
\tfrac{{\mathsf S}_{m \alpha}^2 {\mathsf S}_{m \beta}^2}{\chi(\chi - 1)} \!
\left[ \tfrac{\chi^{-1} - \cos[(\chi - 1)\tau]}{(\chi-1)} \! 
+ \! \tfrac{ \cos (\chi\tau)}{\chi} + 
(1-{ \cos \tau })\!\right]\!. 
\end{equation}
It is interesting to notice that the energy or occupation number of oscillator $a$ 
depends linearly on ${\bar n}_b$.  

In Fig.~\ref{fig3ocn1}, we compare the mean occupation number of oscillator $a$ predicted
by the exact (${\bar n}_a $) and effective models (${\check n}_a) $.  
The latter involving just oscillators $a$, $b$, 
and normal mode $m=1$ ($\Omega=\zeta_1 = \omega$). 
Additionally, we present the oc\-cu\-pa\-ti\-on number of oscillator $b$ using the 
exact model to see how its energy is dynamically depleted to excite oscillator $a$. 
We chose a chain of moderate length 
($N=10$)  in order to be able to progress computationally within the exact model.

%
%%%%%%%%%%%%%%%%%%%%%%%%%%%%%%%%%%%%%%%%%%%%%%%%%%%%%%%%%%%%%%%%%%%%%
\begin{figure}[!bh]                                                                                                                        
\includegraphics[width=8.0cm, trim = 0 20 0 0]{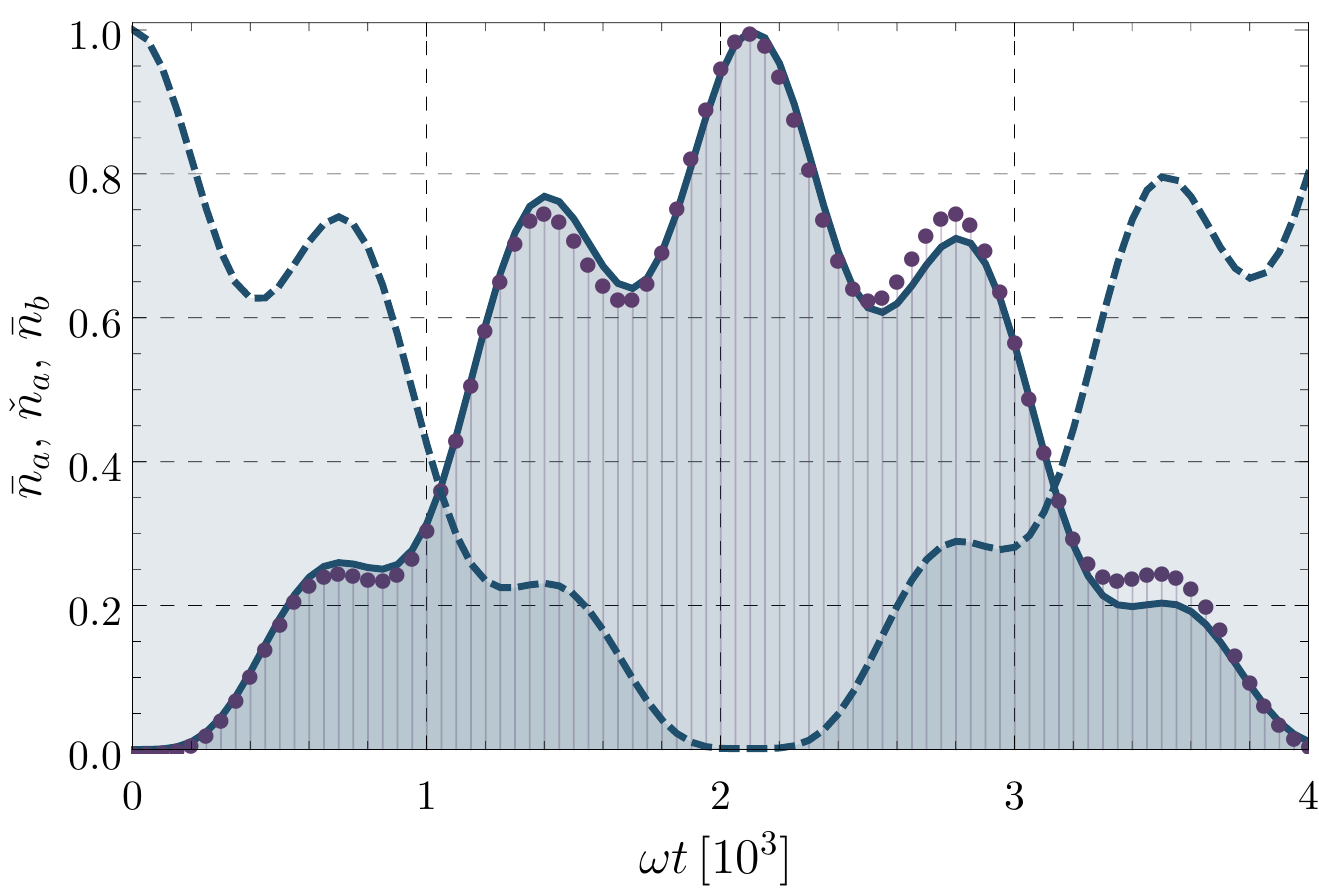}                                                   
\caption{%
Mean occupation number as a function of dimensionless time $\omega t$. 
Solid and dashed lines are exact evolutions for oscillators 
$a$ and $b$, respectively, while dots refer to oscillator $a$ 
using the effective model involving just $a$, $b$ and normal mode $m=1$. 
The chain is composed of $N = 10$ oscillators with first neighbor 
interaction by means o Hooke forces with $\kappa/\omega = 20$. 
Oscillators $a$ and $b$, angular frequency $\Omega = 1$, 
interact with the network also through Hooke forces. 
They couple to network oscillators positioned at 
$\alpha = N$ and $\beta = 1$, respectively (ends of the chain). 
The coupling strength is $\epsilon/\omega = 0.03$.  
Oscillator $b$ starts in a thermal state  
with $\bar{n}_b = 1$, while all other oscillators 
start in local vacuum states.                               
}                                                                                        \label{fig3ocn1}                                                      
\end{figure}                                                          
%%%%%%%%%%%%%%%%%%%%%%%%%%%%%%%%%%%%%%%%%%%%%%%%%%%%%%%%%%%%%%%%%%%%%

We are working in the regime $\epsilon \ll \Omega,\varsigma_k$ ($k = 1,...,N$), 
and it is clear that the simple effective model produces excellent results. Of course, 
as time increases, the agreement is gradually spoiled as a consequence of the fact that 
what supports RWA is a first-order perturbation theory which looses applicability for 
arbitrarily long times. This was previously observed in \cite{plenio2005}. %

In order to deepen our understanding about the order of magnitude of corrections to the 
approximate model, we look more closely to the exact dynamics 
$\bar{n}_a$. 
Especially, the simplified model predicts that the occupation number of 
oscillator $a$ vanishes for ${\bar n}_b = 0$. What does the exact model predict? 
To address this question, in Fig.~\ref{fig4ocn2} we present the time evolution 
of $\bar{n}_a$ for the same physical parameters used in Fig.~\ref{fig3ocn1}, 
except for the initial occupation number of $b$, now taken to be ${\bar n}_b = 0$. 
What we see are high frequency and small amplitude oscillations which contribute little
on average to $\bar{n}_a$.
These contributions coming from small-amplitude fast oscillations are a result of 
terms discarded in the RWA.

%%%%%%%%%%%%%%%%%%%%%%%%%%%%%%%%%%%%%%%%%%%%%%%%%%%%%%%%%%%%%%%%%%%%%
\begin{figure}[!htbp]                                                                                                                        
\includegraphics[width=8.0cm,trim=0 20 0 0]{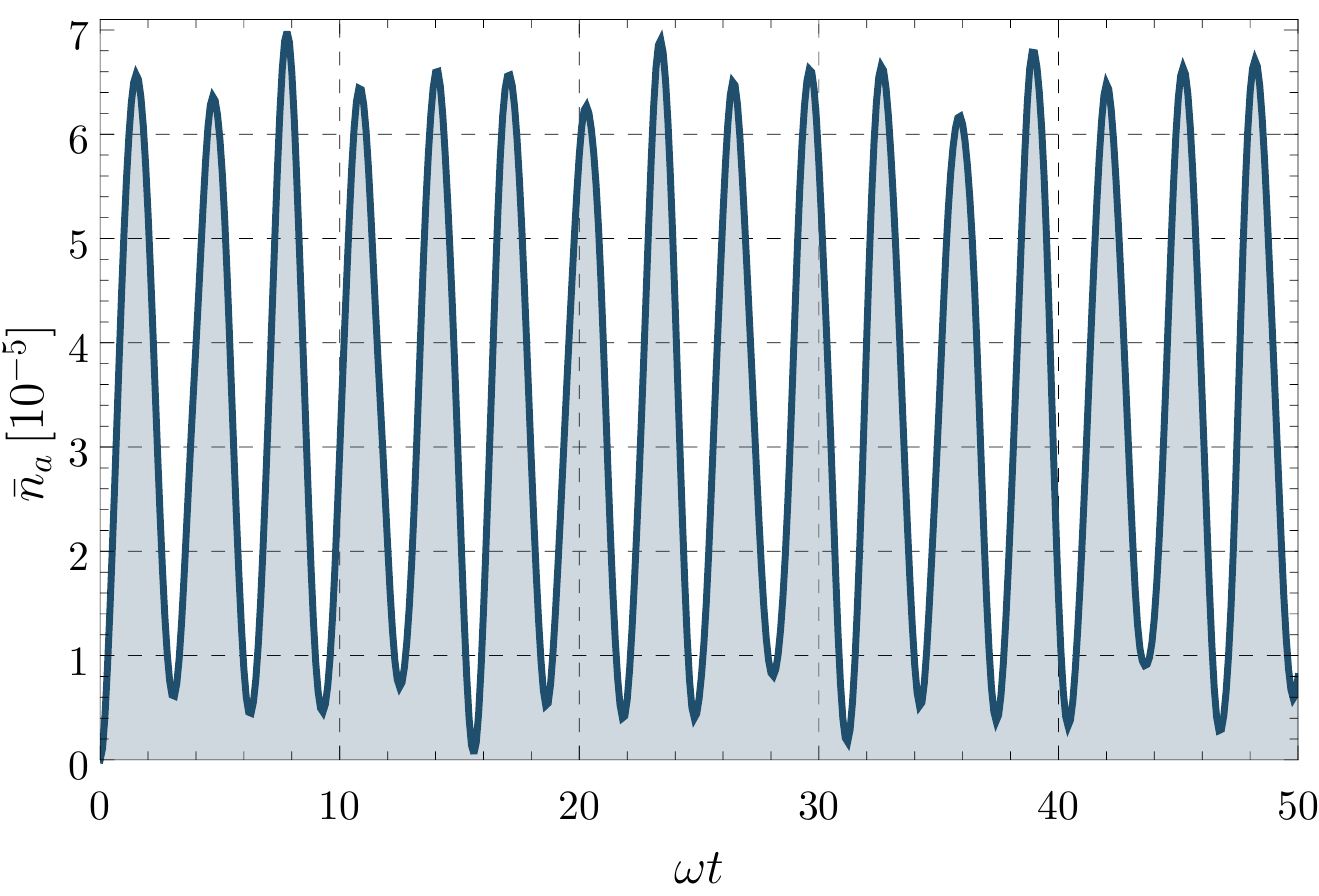}                                                   
\caption{% 
Exact time evolution of the mean occupation number of oscillator $a$ for oscillator 
$b$ initially prepared in a vacuum state ${\bar n}_b = 0$. 
The remaining parameters are kept as in Fig.~\ref{fig3ocn1}.                      
}                                                                                        \label{fig4ocn2}                                                      
\end{figure}                                                          
%%%%%%%%%%%%%%%%%%%%%%%%%%%%%%%%%%%%%%%%%%%%%%%%%%%%%%%%%%%%%%%%%%%%% 

As mentioned before, the effective Hamiltonian changes considerably depending on which 
mode is in resonance with the external oscillators. 
Since the dynamics of (\ref{moneff2}) is entirely determined by the function $F$, 
its analysis should reveal this dependence in a clear way. One can see, for example, 
that for fixed physical parameters ($\epsilon$, ${\bar n}_b$, etc.) and resonance with 
mode $m=1$, the global maximum of $F$, as a function of time, majorates all global 
maxima attained when resonance is set with other modes. Besides, only when resonance 
takes place with $m=1$, the function $F$ is independent of $\alpha$ and $\beta$. 
This rich behavior can be explored for controlling transport 
in the chain \cite{plenio2005}. In order to appreciate this dependence, 
we show in Fig.~\ref{fig5focn}  the dynamical behavior of $F$ for resonance with $m=2$ 
and for $a$ fixed to one of the ends of the chain. One can clearly see the dependence 
on $\beta$, {\it i.e.}, the position in the chain oscillator $b$ is attached to.

%%%%%%%%%%%%%%%%%%%%%%%%%%%%%%%%%%%%%%%%%%%%%%%%%%%%%%%%%%%%%%%%%%%%% 
\begin{figure}[!b]                                                                                                                        
\includegraphics[width=8cm,trim=0 20 0 0]{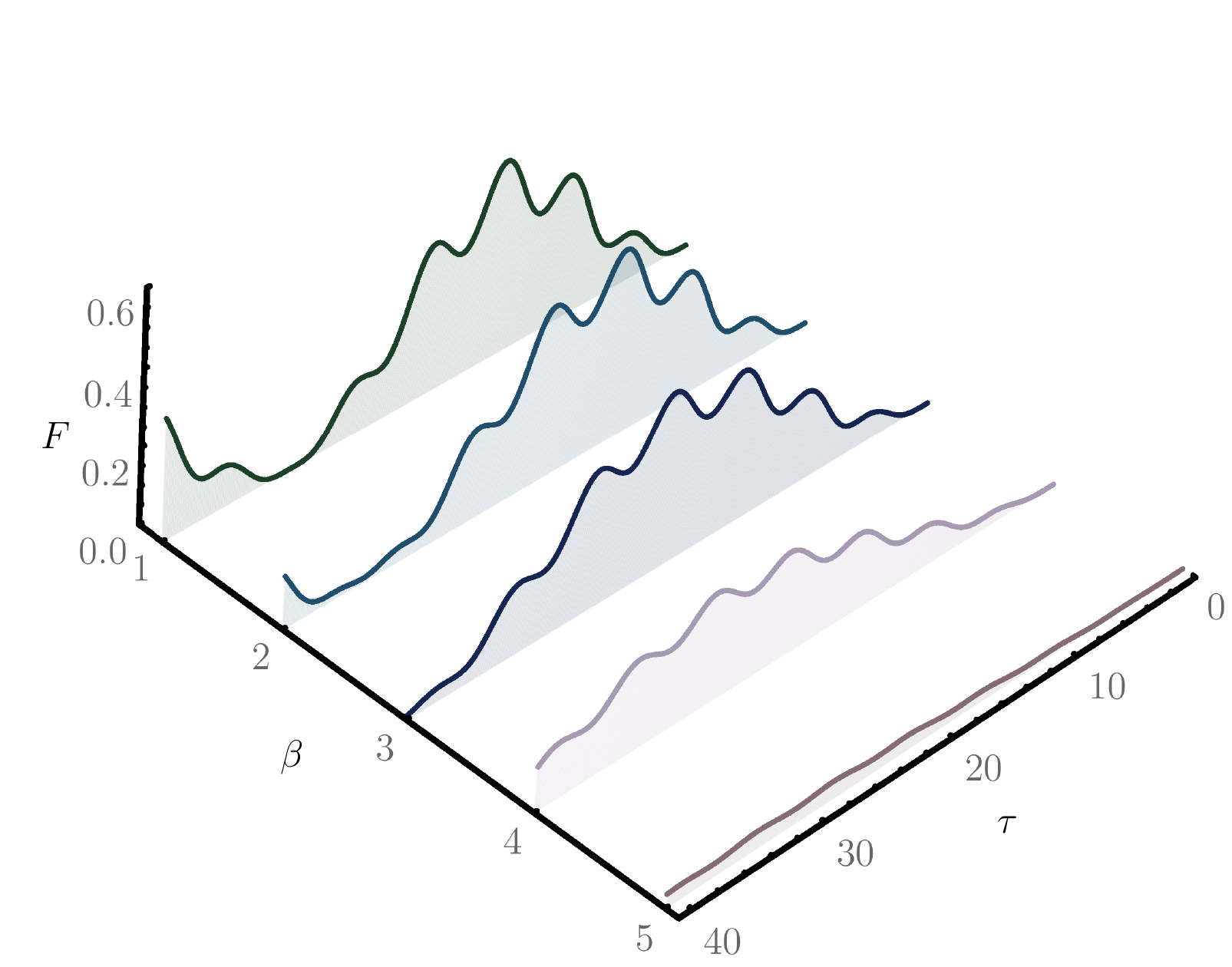}                                                   
\caption{% 
Dependence of function 
$F$ defined in (\ref{funF}) on $\beta$ and 
scaled time $\tau=\epsilon t/4$. 
We consider $\alpha = N = 10$ and $m = 2$. 
The remaining parameters are kept as in Fig.~\ref{fig3ocn1}.                      
}                                                                                        \label{fig5focn}                                                      
\end{figure}                                                          
%%%%%%%%%%%%%%%%%%%%%%%%%%%%%%%%%%%%%%%%%%%%%%%%%%%%%%%%%%%%%%%%%%%%% 

Now, let us suppose that the energy initially in the system is not only due to 
oscillator $b$. For example, the network might as well have some initial thermal energy. 
Would it be possible to theoretically separate the contributions from $b$ and network 
to the energy absorbed by oscillator $a$? To investigate this question, 
we still consider oscillator $b$ initially in a thermal state with thermal 
occupation number $\bar{n}_b$, but now each oscillator in the network is initially 
found in a local thermal state, all of them at same temperature, {\it i.e.}, 
with the same thermal occupation $\bar{n}$. The CM (\ref{cmdef}) for the initial 
global state is then 
\begin{equation}                                                                         \label{globalcm02}
{\bf V}'_0 = 
{\bf V}_{\rm T} \oplus \tfrac{\hbar}{2} (2{\bar n} + 1) \, \mathsf I_{2N}, 
\end{equation} 
with ${\bf V}_{\rm T}$ as in (\ref{cm0}). By using Eq.~(\ref{moneff}) 
and information in Appendix~\ref{ma}, it is tedious but straightforward to show that    
\begin{equation}                                                                         \label{moneff3}
{\check n}'_a = {\check n}_a + 
4 \chi^{-2}  {\mathsf S_{m \beta }^{2} 
\sin^{2}\left( \tfrac{\chi \epsilon t}{8} \right) } \bar n ,
\end{equation}
where ${\check n}_a$ is given in (\ref{moneff2}) and $\chi$ is implicitly defined 
in (\ref{moneff2}) and (\ref{funF}). From this, some comments are in order. 
First of all, one can see that the mean occupation number of  oscillator $a$ is indeed 
the result of distinct contributions from oscillator $b$ and network. 
The latter contributes with the term which does not depend on the occupation number of 
$b$, that is  
$
4 \chi^{-2}  {\mathsf S_{m \beta }^{2} \sin^{2}
\left( \tfrac{\chi \epsilon t}{8} \right) } \bar n
$. 
It is worth noticing that this contribution increases with the temperature of the network 
oscillators as one could expect. The separation between contributions coming from $b$ and
normal mode $m$ is possible because the effective model involves only three bodies and no
direct coupling between $a$ and $b$. Being able to extract this kind of information from 
a complex system is one of the main advantages of simplified but accurate models. 
Another comment concerns the relatively small flux of energy from the network to 
oscillator $a$. Let us consider again the chain with $N=10$ oscillators used to produce 
Fig.~\ref{fig3ocn1}.  Although there were initially a total number of ten thermal phonons 
in the network (one for each of the ten oscillators), only $2.8\%$ of them is 
absorbed by $a$. This can be seen from Fig.~\ref{fig6ocn3} where we show the time 
evolution of (\ref{moneff3}) considering oscillator $b$ in the vacuum state, while the 
ten oscillators of the network are prepared in local thermal states with $\bar{n} =1$. 
The physical explanation for this observation relies on the fact that the initial ten 
thermal photons are shared by all normal modes. When resonance is set to one of these 
modes, the energy in the other modes becomes unavailable to $a$ or $b$.

%%%%%%%%%%%%%%%%%%%%%%%%%%%%%%%%%%%%%%%%%%%%%%%%%%%%%%%%%%%%%%%%%%%%%
\begin{figure}[!b]                                                                                                                        
\includegraphics[width=8.0cm,trim=0 20 0 0]{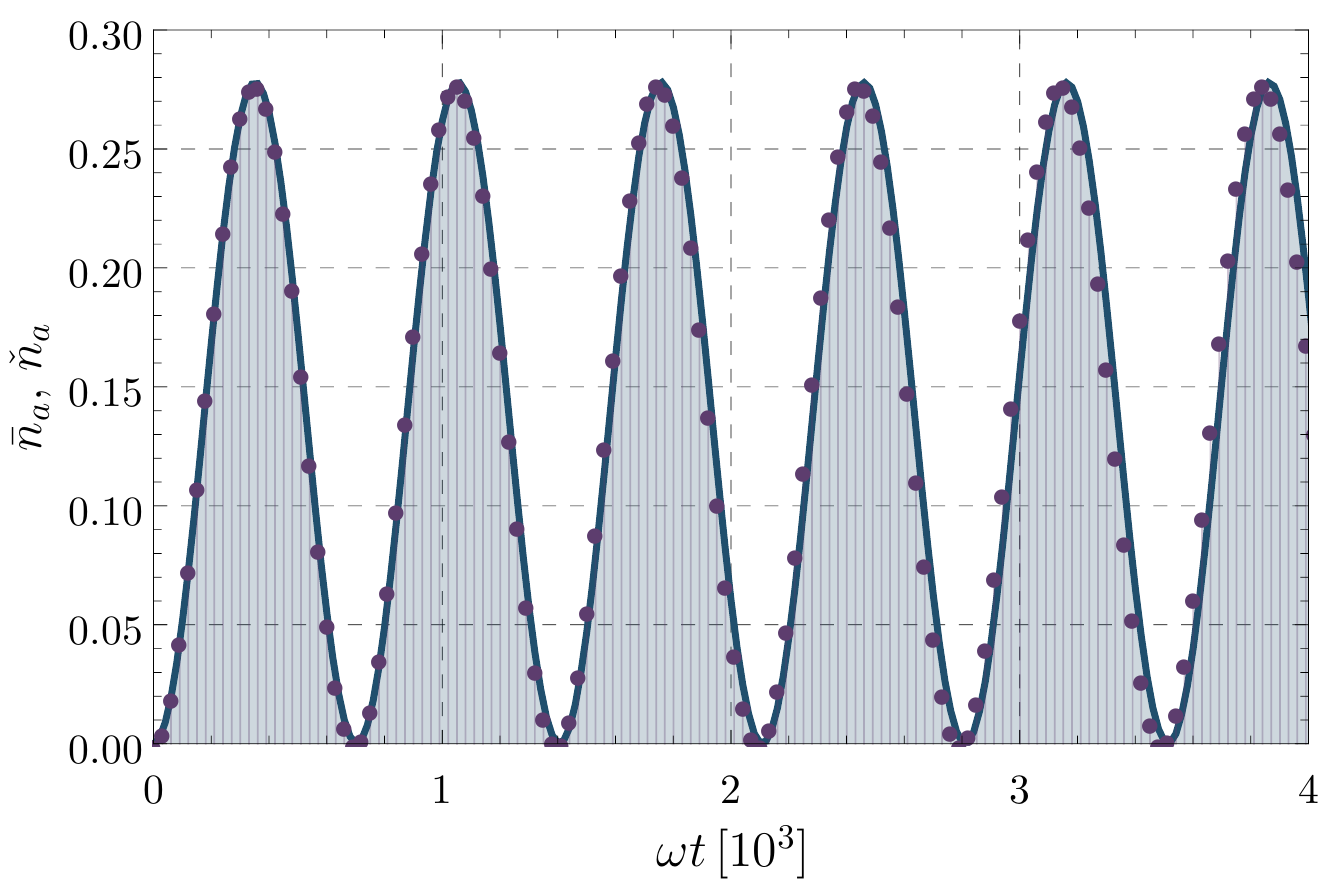}                                                   
\caption{% 
Mean occupation number as a function of dimensionless time $\omega t$. 
The physical parameters are that of Fig.~\ref{fig3ocn1},  
except for the initial state of network oscillators which now 
is the product of local thermal states with $\bar n = 1$, 
and the initial state of $b$ which now is vacuum, 
{\it i.e.}, $\bar n_b = 0$.                              
}                                                                                        \label{fig6ocn3}                                                      
\end{figure}                                                          
%%%%%%%%%%%%%%%%%%%%%%%%%%%%%%%%%%%%%%%%%%%%%%%%%%%%%%%%%%%%%%%%%%%%%

We now move to a scenario where the oscillators (external and network) are 
subjected to local thermal baths whose action on system is due to (\ref{lopthermal}). 
Now, the equation of motion for the CM is given by (\ref{cmev}) with
\begin{equation}                                                                         \label{dynmat1}
{\bf \Gamma} =  - \frac{\zeta}{2} \mathsf I_{2N+4} + \mathsf J \mathbf H,  \,\,\, 
{\bf D }     =    \hbar \zeta({\bar n}_{\text{th}} + \tfrac{1}{2}) \mathsf I_{2N+4},
\end{equation}
and its formal solution is given by (\ref{cmsol}). It is clear that the exponentiation 
of $(2N+4)\times(2N+4)$ causes computational difficulties already 
for moderately high $N$. Besides, it is basically impossible to progress analytically 
within this many body description. 
Using the results developed here, one can give a clear and accurate description of the 
dynamics of the external oscillators working with (\ref{cmsol1}) and (\ref{decmateff2}) 
instead. Now, the matrices to be exponentiated are just $6\times 6$, and one may show 
that 
\begin{equation}                                                                         \label{cmsol4}
\check{\bf V}(t)  = {\rm e}^{-\zeta t} 
                    {\mathsf E}(t) \, \check{\bf V}\!_{0}  \, {\mathsf E}^\top(t) + 
                    \frac{1}{\zeta} 
                    \left( 1 - {\rm e}^{-\zeta t}  \right) \check{\bf D},  
\end{equation}
with ${\mathsf E}(t)$ still given by (\ref{rwssimp2}). 
It is interesting to notice that the unitary part of this evolution, 
already present in (\ref{cmsol3}), is now exponentially attenuated with 
characteristic time $\zeta^{-1}$ in the above equation.

Let us now then compare the predictions using the approximate effective 
model developed here and the exact dynamics. In Fig.~\ref{fig7ocn4}, 
we present the mean occupation number for oscillator $a$ following both descriptions. 
We keep notation used in the closed system case.  Giving the general agreement between 
both descriptions, it is clear that our methodology works 
well also for the open system case. This plot shows that the higher 
the relaxation constant $\zeta$, the sooner the occupation number of 
oscillator $a$ reaches that of the thermal reservoir it is interacting with, 
which is ${\bar n}_{\text{th}}=1$. As time passes, the presence of 
the initial state ${\mathbf V}_0$ 
in (\ref{cmsol4}) is progressively erased by ${\rm e}^{-\zeta t}$, 
which makes the CM tend to 
\begin{equation}                                                                         \label{sstate2}
\lim_{t\to \infty} \check{\bf V}(t)  = \frac{1}{\zeta}\check{\bf D},                 
\end{equation}
showing that not only $a$ thermalizes with its local bath, 
but also $b$ and mode $m$ do the same.

%%%%%%%%%%%%%%%%%%%%%%%%%%%%%%%%%%%%%%%%%%%%%%%%%%%%%%%%%%%%%%%%%%%%%
\begin{figure}[!t]                                                                                                                        
\includegraphics[width=8.0cm,trim=0 20 0 0]{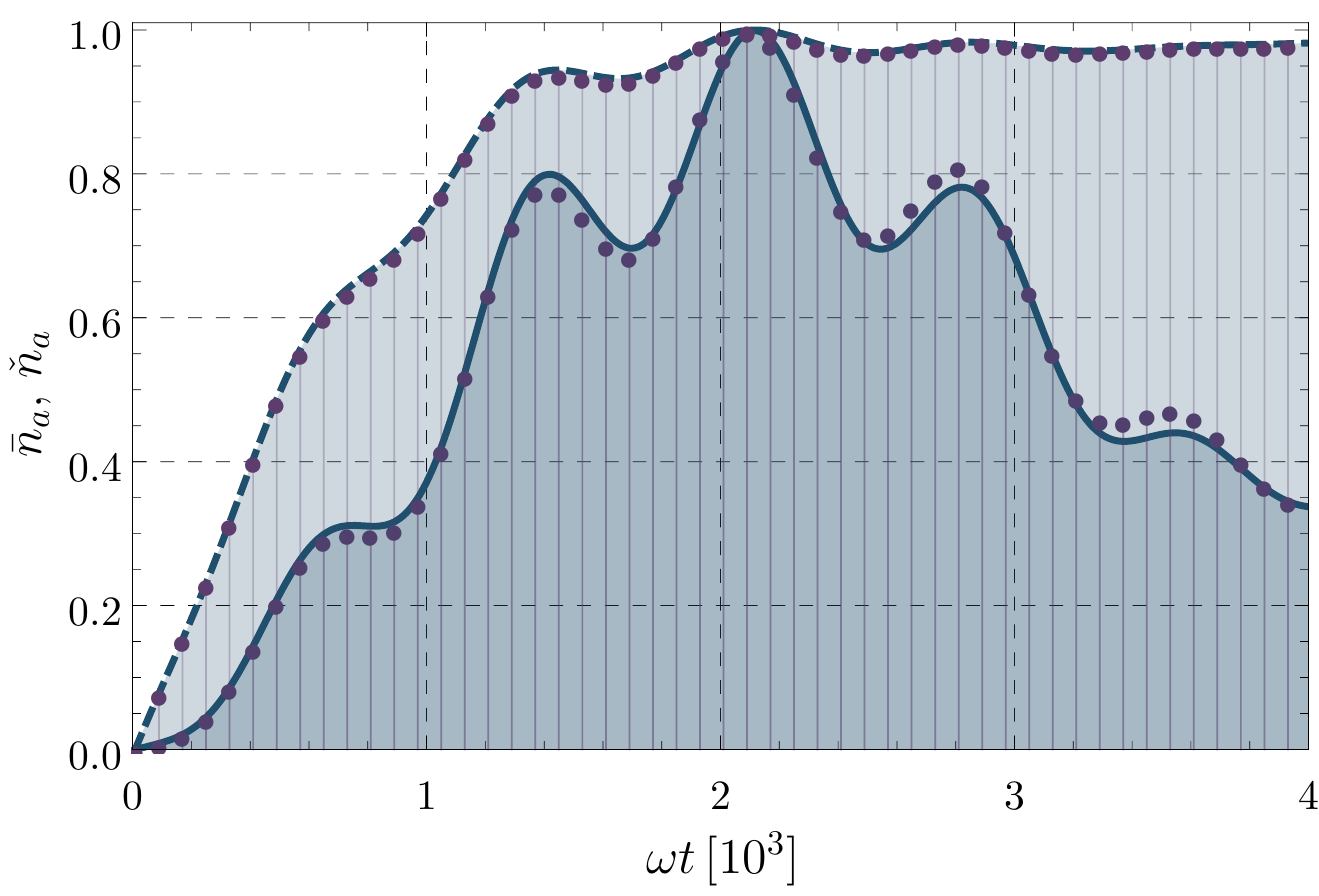}                                                   
\caption{% 
Mean occupation number as a function of dimensionless time $\omega t$. 
Each of the $N+2$ oscillators are attached to 
thermal reservoirs with ${\bar n}_{\text{th}} = 1$. 
Two values of $\zeta$ are considered: $\zeta = 0.01$ (dashed line) and 
$\zeta = 0.001$ (solid line). 
Dots are obtained using the effective model involving just 
$a$, $b$ and normal mode $m=1$. 
The initial state and the remaining parameters are kept the same as in 
Fig.~\ref{fig3ocn1}.                              
}                                                                                        \label{fig7ocn4}                                                     
\end{figure}
%%%%%%%%%%%%%%%%%%%%%%%%%%%%%%%%%%%%%%%%%%%%%%%%%%%%%%%%%%%%%%%%%%%%%

The plot in Fig.~\ref{fig7ocn4} presents another interesting feature. 
The effective model is based on the RWA, 
whose validity is justified with first-order perturbation theory. 
Then, the validity of the approximation is 
limited for finite times. However, we see that the simplified model 
gives the correct asymptotic limit, 
as seen clearly in the case $\zeta = 0.01$. It can be seen with $\zeta = 0.001$ 
as well but at longer times (not shown in the plot).
The reason why the long time regime is not spoiled is that RWA is made only for 
the Hamiltonian part of the dynamics, which becomes less and less important with 
time, see (\ref{cmsol4}) and (\ref{sstate2}).  
The agreement between the complete exact model and 
our simplified model also shows that the decoupling mechanism in terms of  local 
reservoirs in the modes works perfectly well. 
In summary, our model gives very accurate results for the initial cycles of the dynamics 
and for the long time limit when each oscillator is coupled to a thermal bath in the 
conditions discussed here. 
If the system is isolated, {\it i.e.}, no local reservoirs are attached to the oscillators, 
the accuracy will just slowly and gradually be spoiled with time as seen before. 

Now, we want to be more quantitative in terms of the accuracy of the simplified model 
developed here. 
In order to do that, we will investigate the density matrix for oscillator $a$ 
predicted by exact and approximate models, denoted by  $\hat \rho_a$ 
and $\check \rho_a$, 
respectively. We will employ the fidelity $\mathcal{F}$ between these states as a 
figure of merit \cite{scutaru}:
\begin{equation}
\mathcal{F} := \left[ {\rm Tr}
\left(\sqrt{\hat \rho_a} {\check \rho_a}
      \sqrt{\hat \rho_a} \right)^{\frac{1}{2}} \right]^{2} \le 1.
\end{equation}
Since we are working with Gaussian states centered at origin of the phase space, 
one can show that the above formula reduces to \cite{scutaru}
\begin{equation}                                                                         \label{fid1}
\mathcal{F} = \tfrac{2}{ \sqrt{ \det( {\bf V_{\!a}} + \check{\bf V}_{\! \bf a} ) + 
                              \det( {\bf V_{\!a}} -1 )(\check{\bf V}_{\! \bf a} - 1) }  - 
                       \sqrt{\det( {\bf V_{\!a}} -1 )(\check{\bf V}_{\! \bf a} - 1) }  },
\end{equation}
where ${\bf V_{\!a}}$ and $\check{\bf V}_{\! \bf a}$ are, respectively, the CMs of 
subsystem $a$ obtained with (\ref{cmsol2}) and (\ref{cmsol3}) 
\begin{equation}
{\bf V_{\!a}} = 
\begin{pmatrix}
{\bf V}_{11} & {\bf V}_{13} \\
{\bf V}_{31} & {\bf V}_{33}  
\end{pmatrix}, \,\,\, 
\check{\bf V}_{\! \bf a} = 
\begin{pmatrix}
\check{\bf V}_{11} & \check{\bf V}_{13} \\
\check{\bf V}_{31} & \check{\bf V}_{33}  
\end{pmatrix}. 
\end{equation}
%

%%%%%%%%%%%%%%%%%%%%%%%%%%%%%%%%%%%%%%%%%%%%%%%%%%%%%%%%%%%%%%%%%%%%%
\begin{figure}[!t]                                                                                                                        
\includegraphics[width=8.0cm,trim=0 20 0 0]{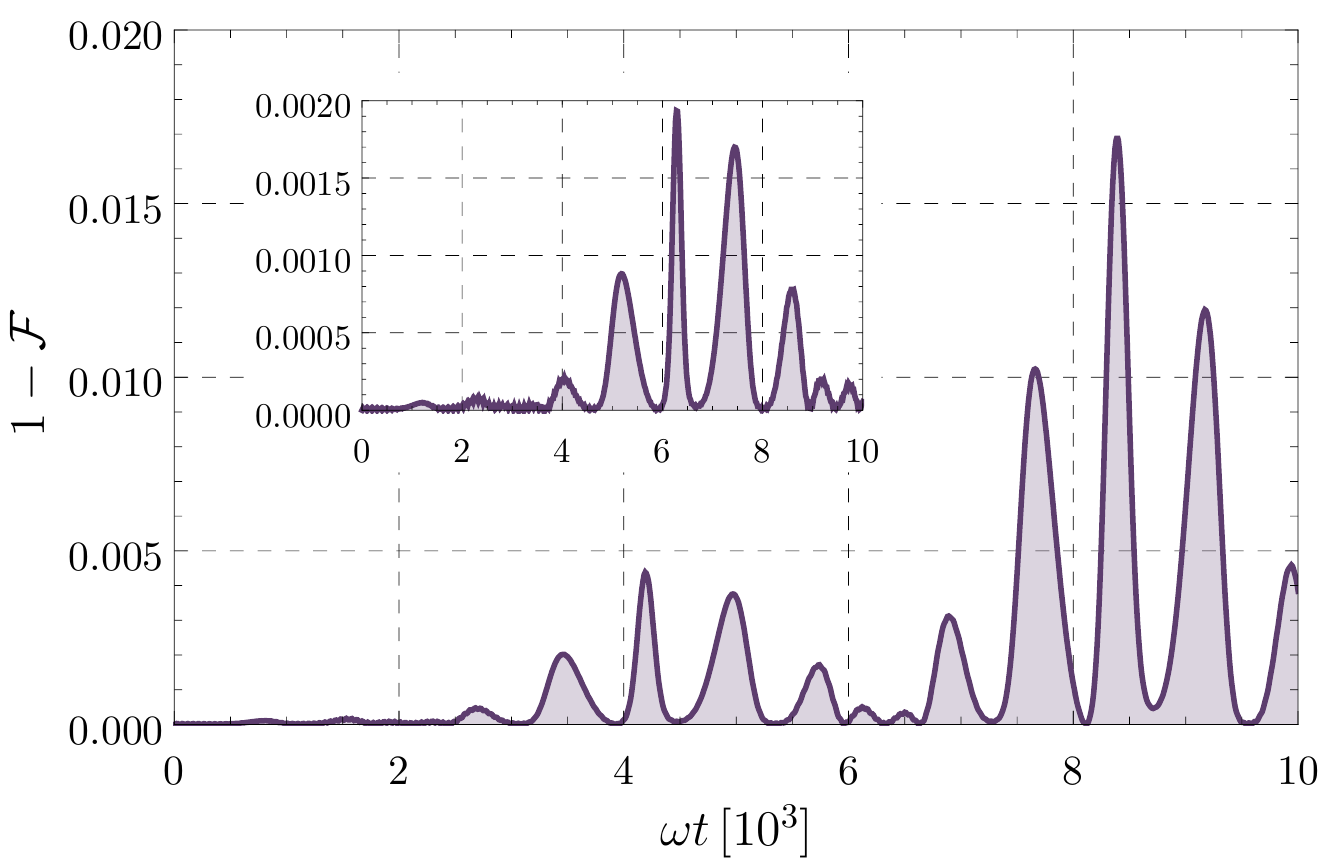}                                                   
\caption{%
Dynamics of $1-\mathcal F$, where $\mathcal F$ is the fidelity 
between evolved density operators for oscillator $a$ using the exact 
and effective models. The physical parameters in the main plot are the 
same as in Fig.~\ref{fig3ocn1}, but in the inset the coupling constant 
with the network is slightly reduced to $\epsilon/\omega = 0.01$.                             
}                                                                                        \label{fig8fid1}                                                      
\end{figure}                                                          
%%%%%%%%%%%%%%%%%%%%%%%%%%%%%%%%%%%%%%%%%%%%%%%%%%%%%%%%%%%%%%%%%%%%%

In Fig.~\ref{fig8fid1}, we plot $(1-\mathcal{F})$ as a function of time for the same 
physical parameters considered in Fig.~\ref{fig3ocn1}. Just like in the plots of 
occupation number, here too the fidelity progressively deteriorates in time, 
which corresponds to the breaking of the RWA for the closed system. However, 
many oscillations are necessary to this deterioration to cause appreciable deviations. 
It is interesting to see that just a slightly reduction of $\epsilon$ 
(inset of Fig.~\ref{fig8fid1}) is enough to make the deterioration even weaker.
This is in complete agreement with the first-order time perturbation theory justification 
of RWA used in this paper.

%%%%%%%%%%%%%%%%%%%%%%%%%%%%%%%%%%%%%%%%%%%%%%%%%%%%%%%%%%%%%%%%%%%%%%%%%%%%%%%%%%%%%%%%%
\subsection{Degenerate normal modes}  \label{d}                       %%%%%%%%%%%%%%%%%%%
%%%%%%%%%%%%%%%%%%%%%%%%%%%%%%%%%%%%%%%%%%%%%%%%%%%%%%%%%%%%%%%%%%%%%%%%%%%%%%%%%%%%%%%%%
To emphasize the generality of effective descriptions based on the methodology developed 
here, let us now consider the system depicted in Fig.~\ref{fig9system3}. 
It opens up the possibility for studying the resonance between the external oscillators 
and degenerate normal modes. 

%%%%%%%%%%%%%%%%%%%%%%%%%%%%%%%%%%%%%%%%%%%%%%%%%%%%%%%%%%%%%%%%%%%%%
\begin{figure}[!htbp]                                                      
\includegraphics[width=8cm,trim=0 0 0 0]{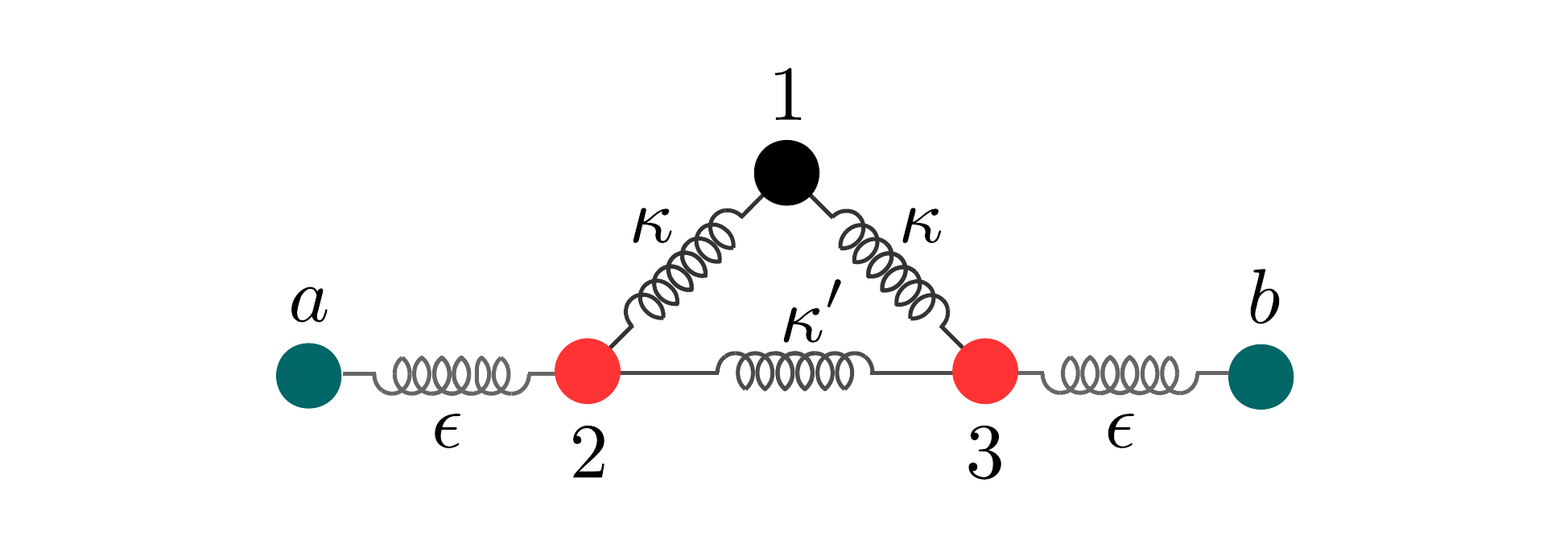} 
\caption{%
Triangular network of oscillators, 
where $\epsilon$, $\kappa$ and $\kappa'$ are coupling constants 
for oscillators coupled by springs (Hooke's forces).
}                                                                                        \label{fig9system3}                            
\end{figure}                                                         
%%%%%%%%%%%%%%%%%%%%%%%%%%%%%%%%%%%%%%%%%%%%%%%%%%%%%%%%%%%%%%%%%%%%% 
  
The free Hamiltonian of the network is written as in Eq.~(\ref{hamfree}) 
but now with 
\begin{equation}                                                                         \label{hesschain2}                                      
{\mathbf H}_{\rm N} = {\bf Q} \oplus {\omega \, \mathsf I_3 }, 
\end{equation}
where the $3 \times 3$ potential matrix is given by
\begin{equation}                                                                         \label{potchain2}
\!{\bf Q}  = \left( 
\begin{array}{ccc}
\kappa + \omega & -\tfrac{1}{2}\kappa &  -\tfrac{1}{2}\kappa \\
-\tfrac{1}{2} \kappa & \tfrac{1}{2}(\kappa + \kappa') + \omega & -\tfrac{1}{2}\kappa' \\
-\tfrac{1}{2}\kappa &  -\tfrac{1}{2}\kappa' & \tfrac{1}{2}(\kappa + \kappa') + \omega
\end{array} \right). 
\end{equation}
The symplectic spectrum (\ref{tw1}) reads now 
\begin{equation}\label{update}
\begin{aligned}
\varsigma_1 &= \omega, \,\,\, \varsigma_2 = \sqrt{\omega(\omega + \kappa/2 + \kappa' )}, 
\,\,\, \\
\varsigma_3 &= \sqrt{\omega(\omega + 3\kappa/2 )}.
\end{aligned}                   
\end{equation}
From (\ref{must}), one can calculate the matrix that performs the 
symplectic diagonalization of the free Hamiltonian. The same structure
as in (\ref{sympdiag}) is found here too, {\it i.e.},
$\mathsf S = {\pmb S} {\pmb O} \! \oplus \! {\pmb S}^{-1} {\pmb O}$,
but now with  
\begin{equation}                                                                         \label{matS}
{\pmb S} = \!
{\rm Diag}\!\left(\! \sqrt[4]{\tfrac{\omega}{\varsigma_1}},\!
\sqrt[4]{\tfrac{\omega}{\varsigma_2}},\!
\sqrt[4]{\tfrac{\omega}{\varsigma_3}} \right)\!, 
{\pmb O} \!=\!\!
\left(\!\!\!\!
\begin{array}{ccc}
\frac{1}{\sqrt{3}}  & \tfrac{1}{\sqrt{3}} & \frac{1}{\sqrt{3}}  \\
0 & - \frac{1}{\sqrt{2}} & \frac{1}{\sqrt{2}} \\
-\frac{\sqrt{2}}{\sqrt{3}} & \frac{1}{\sqrt{6}} & \frac{1}{\sqrt{6}} 
\end{array}\!
\right)\!\!,  
\end{equation}
being $\pmb O$ the orthogonal matrix that performs the Euclidean diagonalization of 
the potential matrix $\bf Q$.  

Considering $\kappa' \neq \kappa $, the effective Hamiltonian is the same 
as in (\ref{hameffinal}) with $\mathbf{H_q}$  (\ref{hesseff2}) defined in terms of 
$\mathsf S_{m \mu}$ calculated using $\mathsf S$ (\ref{matS}) 
with index $m = 1,2,3$ and $\mu = 2,3$. 
The matrices in (\ref{decmateff2}) are also the same, provided we update the symplectic 
eigenvalues to (\ref{update}).   
With this replacement, results 
(\ref{moneff2}), (\ref{moneff3}), (\ref{cmsol4}), and (\ref{sstate2}) stay valid.   

On the other hand, if $\kappa = \kappa'$, the symplectic spectrum is degenerate since 
$\varsigma_2 = \varsigma_3$. As prescribed in Sec.~\ref{eh}, 
if the external oscillators are set in resonance with this degenerate mode, 
$\Omega = \varsigma_2= \varsigma_3$, operator (\ref{degvec}) becomes
$\check x = 
(\hat{\sf q}_a, \hat {\sf q}_b, \hat q_2,\hat q_3,
\hat {\sf p}_a, \hat {\sf p}_b, \hat p_2,\hat p_3)^{\dag}$,            
and the effective dynamics will be governed by (\ref{hameff2}) which, 
for the present case reads 
$\hat H_{\rm eff}^{(2,3)} = 
\tfrac{\epsilon}{8} \check x^\dag \mathbf{H_q}\oplus \mathbf{H_q}\check x$
with 
\begin{equation}                                                                         \label{hesseff3}
\mathbf{H_q} = 
\left( \!\!\! 
\begin{array}{cccc}
1 & 0 & - \mathsf{S}_{ 2 2 } & - \mathsf{S}_{ 3 2} \\
0 & 1 & - \mathsf{S}_{ 2 3 } & -  \mathsf{S}_{ 3 3} \\
- \mathsf{S}_{ 2 2} & - \mathsf{S}_{ 2 3} & 
\mathsf{S}_{  2 2}^2 +  \mathsf{S}_{ 2 3}^2 & 
\mathsf{S}_{ 2 2} \mathsf{S}_{ 3 2} + \mathsf{S}_{ 2 3} \mathsf{S}_{ 3 3}                  \\
- \mathsf{S}_{ 3 2} & - \mathsf{S}_{ 3 3} & 
\mathsf{S}_{ 2 2} \mathsf{S}_{ 3 2} + \mathsf{S}_{ 2 3} \mathsf{S}_{ 3 3} & 
\mathsf{S}_{  3 2}^2 +  \mathsf{S}_{ 3 3}^2 
\end{array}\!\!\!\!\right),
\end{equation}
to be evaluated with (\ref{matS}).
Now one can calculate 
$\mathsf E (t) = \exp\left[ \mathsf J  {\mathbf H}_{\rm eff}\, t  \right] 
\in {\rm Sp}(8,\mathbb R)$
and determine the dynamics of the occupation number for oscillator $a$ which is 
plotted in 
Fig.~\ref{fig10ocn5}. 

%%%%%%%%%%%%%%%%%%%%%%%%%%%%%%%%%%%%%%%%%%%%%%%%%%%%%%%%%%%%%%%%%%%%%
\begin{figure}[!htbp]                                                                                                                        
\includegraphics[width=8.0cm,trim=0 20 0 0]{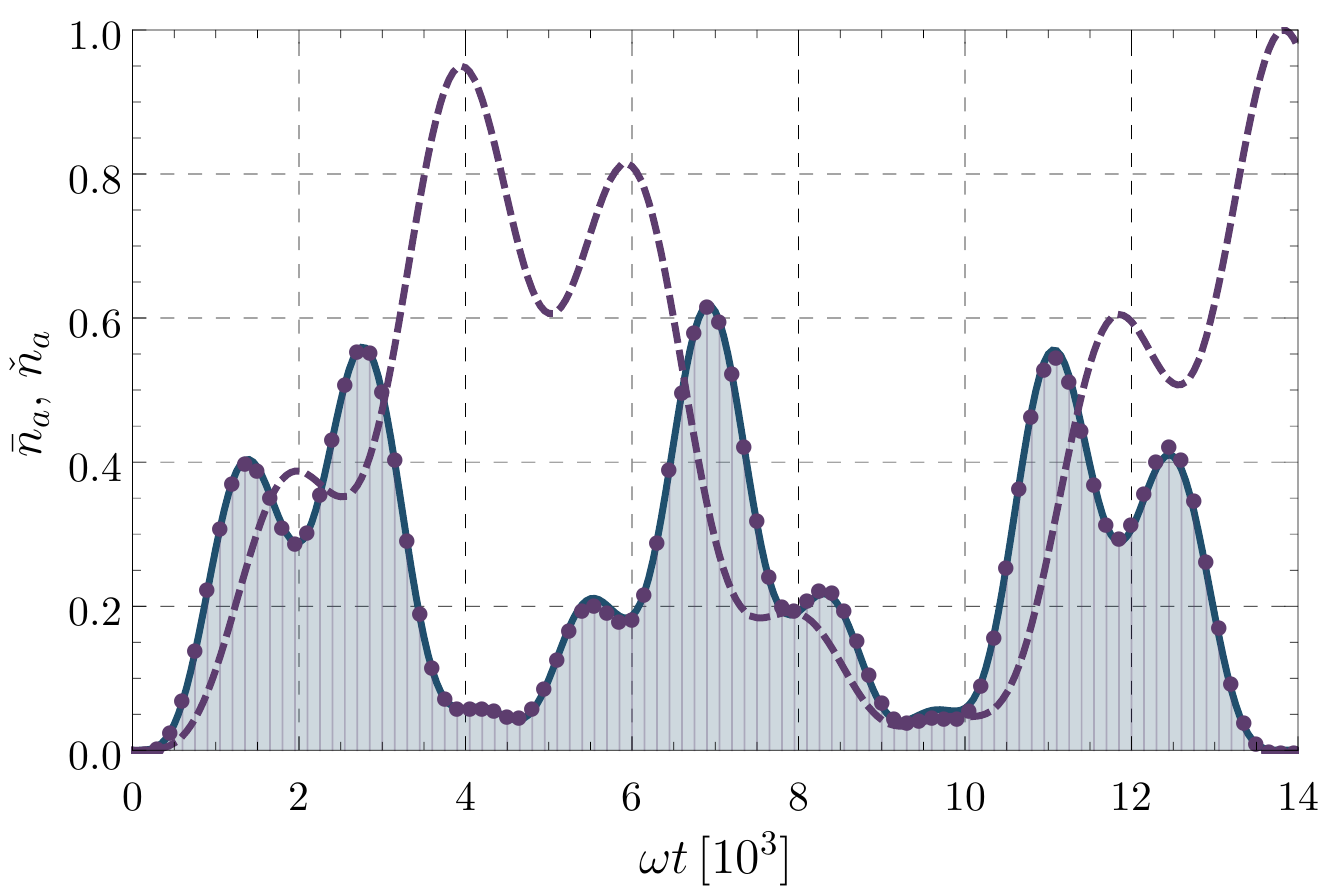}                                                   
\caption{% 
Mean occupation number as a function of dimensionless time $\omega t$. 
The topology depicted Fig.~\ref{fig9system3} is used with 
$\kappa = \kappa'$, and resonance is taken with the resulting 
degenerate modes $\Omega = \varsigma_2=\varsigma_3$.
As before, the solid line is exact and dots refer to the 
approximate model. 
The dashed line corresponds to the dynamics where one mistakably 
and naively includes only mode $3$ in the effective model. 
We consider 
$\kappa/\omega = \kappa'/\omega = 1/3$ and
$\epsilon/\omega = 1/600$. The initial state of oscillator $b$ 
is thermal with $\bar{n}_b = 1$, and the other oscillators 
are in local vacuum states.        
}                                                                                        \label{fig10ocn5}                                                      
\end{figure}                                                          
%%%%%%%%%%%%%%%%%%%%%%%%%%%%%%%%%%%%%%%%%%%%%%%%%%%%%%%%%%%%%%%%%%%%%

Again, it is remarkable the agreement of the simplified model (now two-mode) with exact 
dynamics. For comparison, it is also shown the behavior with just one mode 
in the effective description. 
The reason it to draw our attention to the fact that degeneracy should be taken into 
account carefully through the effective description (\ref{hameff2}). 
For longer times, not shown in the plot, the mean occupation number 
$\bar n_a$ attain $\bar n_b = 1$ within the precision of the numerical treatment of 
the original model. The effective model can not be used at such long times as previously 
discussed.

The inclusion of thermal baths to each oscillator is made along the lines of the previous 
examples (\ref{decmateff2}). Now, one should only be careful to take into account 
the presence of one more 
mode, {\it i.e.},
\begin{equation}                                                                         \label{decmateff3} 
\begin{aligned}                                                                
\check{\bf \Gamma } &:= 
\mathsf J^{[8]} \left( \mathbf{H_q}  \oplus \mathbf{H_q}  \right)
 - \frac{\zeta}{2} \mathsf I_{8} , \\                                                                           
 \check{\bf D} & :=  \hbar \zeta({\bar n}_{\text{th}} + \tfrac{1}{2}) 
\left( \mathsf I_2 \oplus \tfrac{\varsigma_2}{\omega} 
\oplus \tfrac{\varsigma_3}{\omega} \oplus 
\mathsf I_2 \oplus \tfrac{\omega}{\varsigma_2} 
\oplus \tfrac{\omega}{\varsigma_3}    \right).
\end{aligned}
\end{equation}
The effect is essentially the same as in Fig.~\ref{fig7ocn4} and, for this reason, 
we will not add a plot for this case.

Control of errors due to the approximations made to obtain (\ref{hesseff3}) 
is again made through inspection of $(1-\mathcal{F})$, with $\mathcal F$ defined 
in (\ref{fid1}). This is presented in Fig.~\ref{fig11fid2}. One can see that, 
in agreement to what is shown in Fig.~\ref{fig10ocn5}, fidelility is quite 
high meaning that the effective model produces accurate results even in the case of 
degeneracy. As expected, the fidelity slowly degrades with time.
%
%%%%%%%%%%%%%%%%%%%%%%%%%%%%%%%%%%%%%%%%%%%%%%%%%%%%%%%%%%%%%%%%%%%%%
\begin{figure}[!htbp]                                                                                                                        
\includegraphics[width=8.0cm,trim=0 20 0 0]{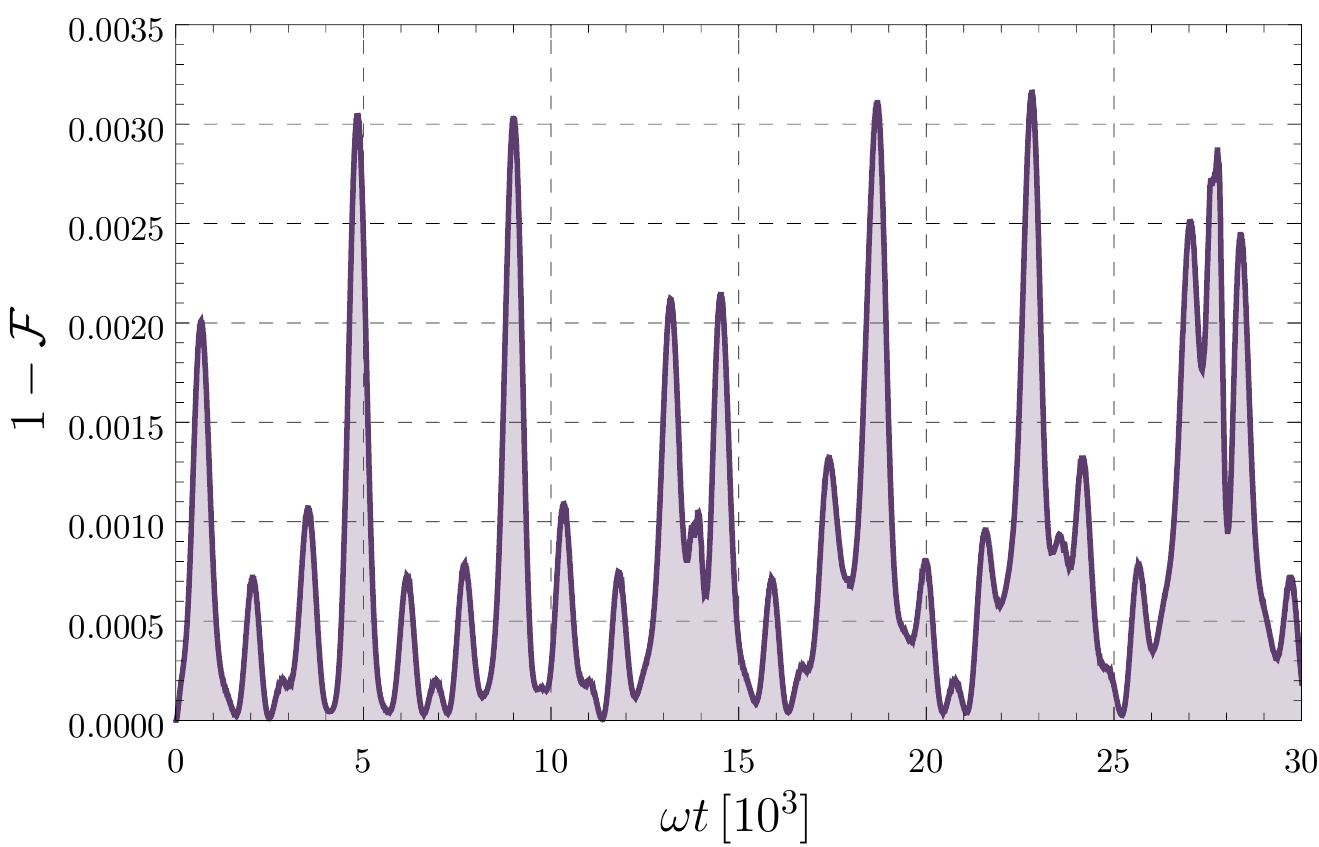}                                                   
\caption{% 
Dynamics of $1-\mathcal F$, where $\mathcal F$ is the fidelity 
between evolved density operators for oscillator $a$ using 
the exact and effective models. This is a case with degeneracy 
and the parameters are those considered in Fig.~\ref{fig10ocn5}.                              
}                                                                                        \label{fig11fid2}                                                      
\end{figure}                                                          
%%%%%%%%%%%%%%%%%%%%%%%%%%%%%%%%%%%%%%%%%%%%%%%%%%%%%%%%%%%%%%%%%%%%%

%%%%%%%%%%%%%%%%%%%%%%%%%%%%%%%%%%%%%%%%%%%%%%%%%%%%%%%%%%%%%%%%%%%%%%%%%%%%%%%%%%%%%%%%%
\subsection{Beyond the Hooke's Law}                                   %%%%%%%%%%%%%%%%%%%
%%%%%%%%%%%%%%%%%%%%%%%%%%%%%%%%%%%%%%%%%%%%%%%%%%%%%%%%%%%%%%%%%%%%%%%%%%%%%%%%%%%%%%%%%
In previous examples, the interaction between oscillators in the network follows Hooke's 
law, {\it i.e.}, spring-like couplings. 
This implies that the effective Hamiltonian in (\ref{hesseff}) does not present crossed 
terms involving position and momentum. Mathematically, this is the same as 
${\bf C_{\bf qp}}$ null in (\ref{hesseffb}).
Since the method is applicable to any positive definite Hamiltonian, 
this section advances to the consideration of a toy model where momentum 
and position cross in the interaction Hamiltonian. For this purpose, 
we consider now that (\ref{hamfree}) is defined with
\begin{equation}                                                                         \label{hammp}
{\bf H}_{\rm N} =   
\left(\begin{array}{cc}
      \omega \mathsf I_{3} & {\mathbf C} \\
      {\mathbf C}          & \omega \mathsf I_{3}
      \end{array}
\right),  \,\,\, 
{\bf H}_{\rm e} = \Omega \mathsf I_4,
\end{equation}
where we considered $N = 3$ oscillators in the network and 
\begin{equation}
\mathbf C = \frac{\gamma}{2} \, \mathsf I_3 - \frac{1}{\sqrt{2}}
\left(\begin{array}{ccc}
      0 & \kappa & 0  \\
      \kappa & 0 & \kappa \\
      0 & \kappa & 0 
\end{array}
\right). 
\end{equation}
The above matrix will lead to 
$- \frac{\kappa}{\sqrt{2}} (\hat q_1\hat p_2 + \hat q_2\hat p_1 + 
                            \hat q_3\hat p_2 + \hat q_2 \hat p_3)
+ \gamma\sum_{j=1}^{3}(\hat q_i\hat p_i + \hat p_i\hat q_i) $
 in the network Hamiltonian.
Since ${\bf H}_{\rm N}$ must be positive definite, 
condition $\omega > \kappa + \gamma$ has to be imposed. 
The external oscillators, $a$ and $b$, interact with the network as usual, 
see (\ref{hamint}).   

For this example, symplectic diagonalization of ${\bf H}_{\rm N}$ results in
\begin{equation}
\begin{aligned}
\varsigma_1 & = \sqrt{\omega^2 - (\kappa+\gamma)^2},\,\,\, 
\varsigma_2 = \sqrt{\omega^{2} - (\kappa-\gamma)^{2}}   ,  \\
\varsigma_3 & =  \sqrt{\omega^{2} - \gamma^{2}} ,
\end{aligned}                                
\end{equation}
from which one obtains the matrix that performs the 
symplectic diagonalization of  ${\bf H}_{\rm N}$. This is now written as 
$\mathsf S_{\rm N} = ({\pmb S} \oplus {\pmb S}^{-1}) \mathsf R ({\pmb O} \oplus {\pmb O})$,
where
\begin{equation} 
\begin{aligned}                                                                        \label{matS2}
{\pmb S} & = 
{\rm Diag}\left( \sqrt[4]{\tfrac{\omega - \kappa - \gamma}{\omega + \kappa + \gamma}}, 
                 \sqrt[4]{\tfrac{\omega + \kappa - \gamma}{\omega - \kappa + \gamma}},
                 \sqrt[4]{\tfrac{\omega - \gamma}{\omega + \gamma}}   \right),  \\
\mathsf R & = \frac{1}{\sqrt{2}}
\left(
\begin{array}{cc}
 \mathsf I_3 & \mathsf I_3  \\
-\mathsf I_3 & \mathsf I_3
\end{array}
\right), \,\,\, \\
{\pmb O} & =
\left(
\begin{array}{ccc}
\frac{1}{2}  & -\frac{1}{\sqrt{2}} & \frac{1}{2}  \\
\frac{1}{2} & \frac{1}{\sqrt{2}} & \frac{1}{2} \\
-\frac{1}{\sqrt{2}} & 0 & \frac{1}{\sqrt{2}} 
\end{array}
\right),  
\end{aligned}
\end{equation}
being $\pmb O$ the orthogonal matrix that performs Euclidean diagonalization of $\bf C$. 
Writing the effective Hamiltonian (\ref{hesseff}) for $\Omega = \varsigma_1$, 
we can work on the time evolution of the covariance matrix (\ref{cmsol2}) 
to obtain the mean occupation number of the oscillator $a$, 
as plotted in Fig.~\ref{fig12ocn6}. 
Again, the effective model agrees quite well with the exact dynamics. 
The inclusion of local thermal baths would follow just like 
before, since again $\mathsf S_{\rm N}\mathsf S_{\rm N}^\top$ is a diagonal matrix.

%%%%%%%%%%%%%%%%%%%%%%%%%%%%%%%%%%%%%%%%%%%%%%%%%%%%%%%%%%%%%%%%%%%%%
\begin{figure}[!tp]                                                                                                                        
\includegraphics[width=8.0cm,trim=0 20 0 0]{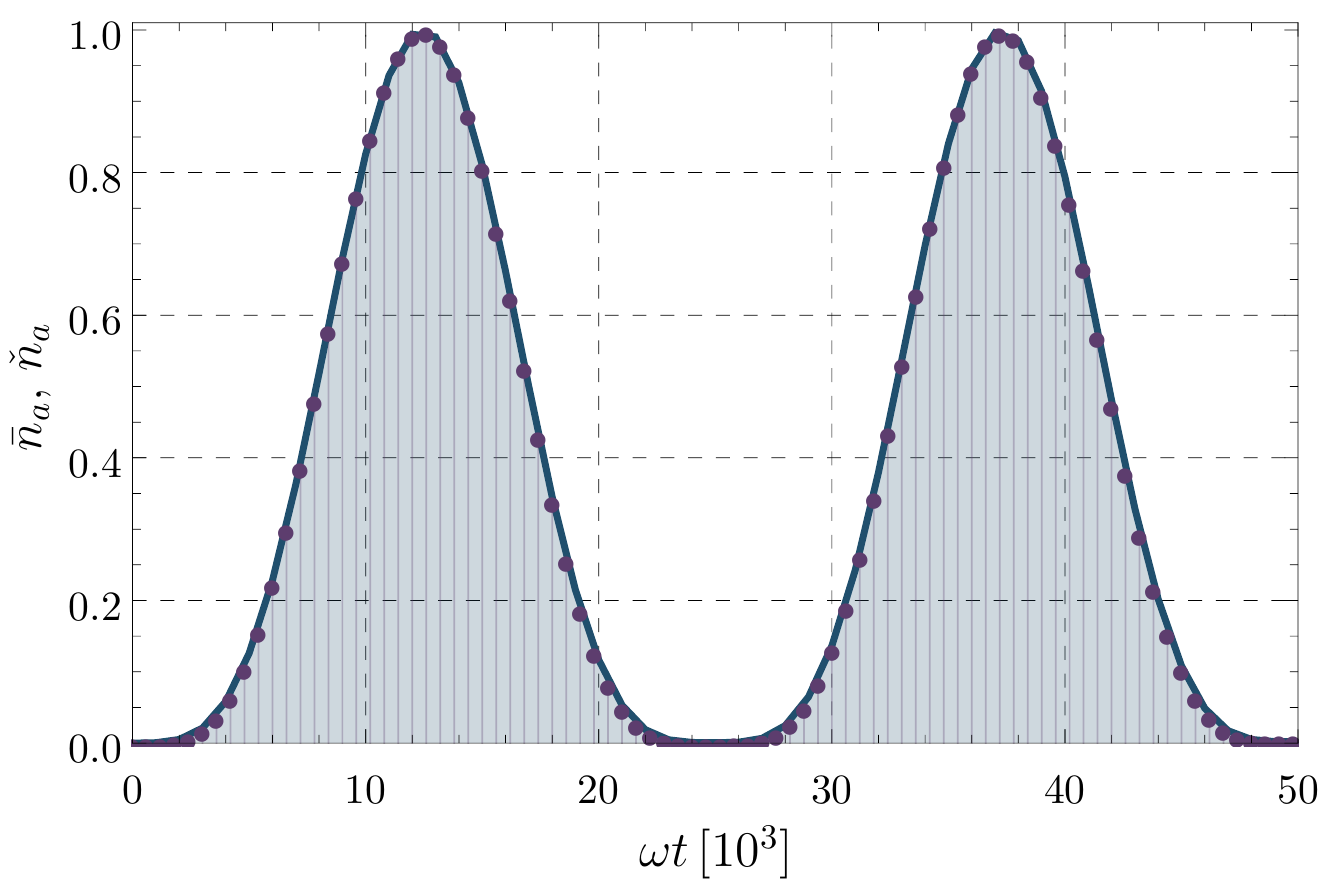}                                                   
\caption{%
Mean occupation number of oscillator $a$ as a function of
dimensionless time $\omega t$.  The Hamiltonian of the network 
is given by (\ref{hammp}) with 
$\kappa/\omega = 0.5$ and $\gamma/\omega = 0.2$. 
The solid line is the exact time evolution, 
while the dots are the result of the effective model. Oscillators $a$ and 
$b$ possess frequency $\Omega = \varsigma_1$ and are coupled, respectively, 
to oscillators $\alpha = 1$ and $\beta = 3$ in the network with 
$\epsilon/\omega = 0.001$. %
Oscillator $b$ is initially in a thermal state 
 with $\bar{n}_b = 1$ and 
all other oscillators are initially in local vacuum states.                               
}                                                                                        \label{fig12ocn6}                                                      
\end{figure}                                                          
%%%%%%%%%%%%%%%%%%%%%%%%%%%%%%%%%%%%%%%%%%%%%%%%%%%%%%%%%%%%%%%%%%%%%

Fidelity is again used to infer the quality of the approximations made to obtain the 
simplified model, see Fig.~\ref{fig13fid3}. 
The result shows that the accuracy of the effective description is again remarkable.
%
%%%%%%%%%%%%%%%%%%%%%%%%%%%%%%%%%%%%%%%%%%%%%%%%%%%%%%%%%%%%%%%%%%%%%
\begin{figure}[!htbp]                                                                                                                        
\includegraphics[width=8.0cm,trim=0 20 0 0]{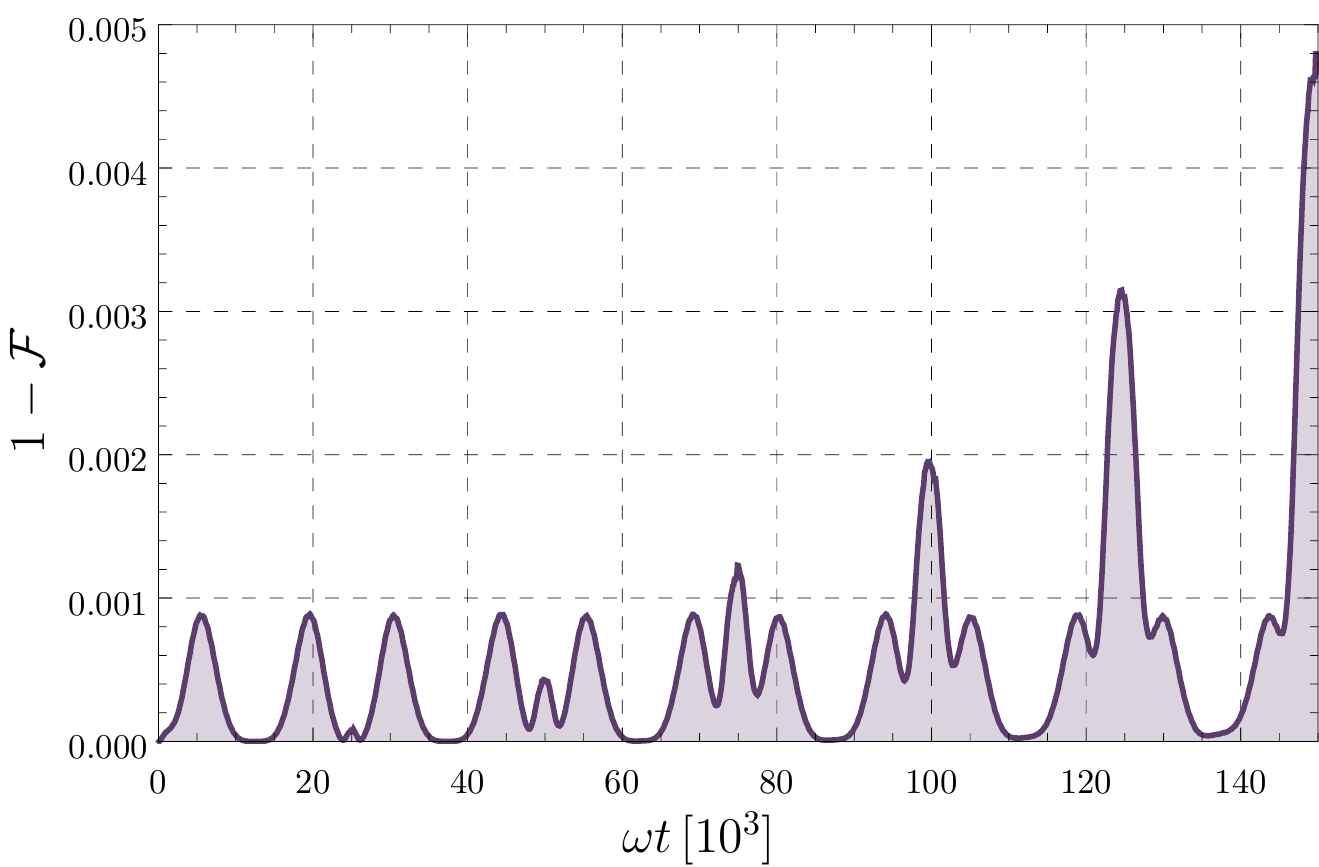}                                                   
\caption{% 
Dynamics of $1-\mathcal F$, where $\mathcal F$ is the fidelity between evolved density 
operators for oscillator $a$ using the exact and effective models. 
This plot refers to the case considered in Fig.~\ref{fig12ocn6}.                              
}                                                                                        \label{fig13fid3}                                                      
\end{figure}                                                          
%%%%%%%%%%%%%%%%%%%%%%%%%%%%%%%%%%%%%%%%%%%%%%%%%%%%%%%%%%%%%%%%%%%%%  

%%%%%%%%%%%%%%%%%%%%%%%%%%%%%%%%%%%%%%%%%%%%%%%%%%%%%%%%%%%%%%%%%%%%%%%%%%%%%%%%%%%%%%%%%
\section{ Final Remarks } \label{fr}                                  %%%%%%%%%%%%%%%%%%%
%%%%%%%%%%%%%%%%%%%%%%%%%%%%%%%%%%%%%%%%%%%%%%%%%%%%%%%%%%%%%%%%%%%%%%%%%%%%%%%%%%%%%%%%%
% 
We have described a general method to obtain useful and accurate effective descriptions 
of large open systems formed by coupled harmonic oscillators. The idea is that two 
external oscillators weakly coupled to a network of harmonic oscillators may have 
their dynamics 
effectively described by a model with just a few coupled degrees of freedom. 
This was first seen for the linear case of first neighbor coupled oscillators 
with no degeneracy nor thermal reservoirs in \cite{plenio2005}.  
We improve and expand this idea by considering any topology of the network and 
by including environments for all elements of the system. For the unitary case, 
the only restriction is that the system Hamiltonian must be positive definite. 
When environments are attached to the oscillators, we show that further structural 
restrictions must be imposed to grant the simplified descriptions. In general, 
we showed that the number of effectively coupled constituents depends on the nature 
of the symplectic spectrum of the Hessian of the Hamiltonian and resonances.

As an application and illustration of the method, we consider the problem of 
propagation of energy through the network. Meaningful and informative analytical 
results could be obtained in the scope of the simplified model. 
We also presented how fidelity between the evolved states under exact and effective 
descriptions behaves, and the result shows that the accuracy of the simplified 
model is quite remarkable. Different topologies are used to illustrate the 
applicability of the methodology presented here. 

It is worthwhile noticing that instead of coupling single harmonic oscillators 
to the network, one could have networks coupled to networks and obtain simplified 
models involving few coupled normal modes of different networks. 
In this case, the first step is symplectic diagonalization of each network and then, 
through resonances, end up with an effective model following our recipe.

Our work considers finite networks as environments for the two external oscillators in 
contrast to the typical baths necessary to model a system-reservoir interaction. 
In the latter, the network is formed by an infinity set of harmonic oscillators 
or continuum spectrum of modes. Our approach is of interest because it applies 
to this intermediate case where the network is big enough for not being amenable to 
analytical exact treatment and, at the same time, it is not big enough for allowing 
the usual approximations that follow from interaction with a bath. 
These approximations are needed, in general, for ending up with a useful master equation.}  

It is our opinion that the present study can contribute to studies involving 
transport of different physical resources in coupled harmonic systems by 
allowing effective descriptions amenable to analytical progress. 
This is important to the evaluation, for example, of limits in channel capacities 
or even stationary heat currents, just to name a few direct applications.

During the the revision of this work, we became aware of  \cite{galve} which treats a 
similar system but from a different point of view whose approach is subjected to different 
limits of validity and aims than ours.  

%%%%%%%%%%%%%%%%%%%%%%%%%%%%%%%%%%%%%%%%%%%%%%%%%%%%%%%%%%%%%%%%%%%%%%%%%%%%%%%%%%%%%%%%%
\acknowledgments                                                      %%%%%%%%%%%%%%%%%%%
%%%%%%%%%%%%%%%%%%%%%%%%%%%%%%%%%%%%%%%%%%%%%%%%%%%%%%%%%%%%%%%%%%%%%%%%%%%%%%%%%%%%%%%%%
FN and FLS are supported by the CNPq ``Ci\^{e}ncia sem Fronteiras'' 
programme through the ``Pesquisador Visitante Especial'' initiative 
(grant nr. 401265/2012-9).
FLS is a member of the Brazilian National Institute of Science and Technology of 
Quantum Information (INCT-IQ) and acknowledges partial support from CNPq 
(grant nr. 307774/2014-7).

%%%%%%%%%%%%%%%%%%%%%%%%%%%%%%%%%%%%%%%%%%%%%%%%%%%%%%%%%%%%%%%%%%%%%%%%%%%%%%%%%%%%%%%%%
\section*{ Appendix } \appendix                                       %%%%%%%%%%%%%%%%%%%
%%%%%%%%%%%%%%%%%%%%%%%%%%%%%%%%%%%%%%%%%%%%%%%%%%%%%%%%%%%%%%%%%%%%%%%%%%%%%%%%%%%%%%%%%
%
%%%%%%%%%%%%%%%%%%%%%%%%%%%%%%%%%%%%%%%%%%%%%%%%%%%%%%%%%%%%%%%%%%%%%%%%%%%%%%%%%%%%%%%%%
\section{ Rotating Wave Approximation } \label{rwa}                   %%%%%%%%%%%%%%%%%%%
\renewcommand{\theequation}{A-\arabic{equation}}                      %%%%%%%%%%%%%%%%%%%
%%%%%%%%%%%%%%%%%%%%%%%%%%%%%%%%%%%%%%%%%%%%%%%%%%%%%%%%%%%%%%%%%%%%%%%%%%%%%%%%%%%%%%%%%
Consider two interacting harmonic oscillators (frequencies $\omega_1$ and $\omega_2$) 
whose dynamics is ruled by Hamiltonian $\hat{H}=\hat H_0+{\hat H}_I$, where 
\begin{equation}                                                                         \label{hamsimp}
\begin{aligned}
&\hat H_0 = \hbar \omega_1 \hat a_1^{\dagger} \hat a_1 + 
            \hbar \omega_2 \hat a_2^{\dagger} \hat a_2,\\
&{\hat H}_I = \hbar\sum_{j\neq k}
                {\eta}_{jk} {\hat a}_j^{\dagger} {\hat a}_k + \hbar\sum_{j,k }
                {\xi}_{jk} {\hat a}_j {\hat a}_k +  
                {\xi}_{jk}^{\ast} {\hat a}_j^{\dagger}{\hat a}_k^{\dagger},  
\end{aligned}
\end{equation}
$\eta_{jk}$ and $\xi_{jk}$ time independent arbitrary numbers,  
$\hat a_j$ is the annihilation operator of the oscillator $j$ 
and $i,j=1,2$.   
With respect to $\hat H_0$, the interaction picture Hamiltonian reads
\begin{equation}\label{in}
{\tilde H}_I = \hbar\sum_{j\neq k}
                {\eta}_{jk} {\tilde a}_j^{\dagger} {\tilde a}_k + \hbar\sum_{j,k}
                {\xi}_{jk} {\tilde a}_j {\tilde a}_k +  
                {\xi}_{jk}^{\ast} {\tilde a}_j^{\dagger}{\tilde a}_k^{\dagger},  \
\end{equation}
where ${\tilde a}_j = {\rm e}^{-i \omega_j t} {\hat a}_j$ and $\hat a_j$.
Since the free Hamiltonian $\hat H_0$ is that of non-interacting oscillators, 
its eigenvectors are just the tensor product of 
Fock  states {\it i.e.}, 
$ %\begin{equation}                                                                      \label{psi1}
| \Psi \rangle = |\mathfrak{n}_1,\mathfrak{n}_2 \rangle, 
$ %\end{equation}
where $ \{ | \mathfrak{n}_j \rangle ; j=1,2 \} $ 
are the eigenstates of 
$ \hat{a}^{\dag}_j \hat{a}_j $ . 

Treating  ${\hat H}_I $ as a perturbation to $\hat H_0$, first order time dependent 
perturbation theory reveals that the transition probability between two eigenstates 
of $\hat H_0$, denoted $| \Psi \rangle$ and $| \Psi' \rangle$, is given by \cite{cohen}
\begin{equation}                                                                         \label{tp}
\mathcal P_{\Psi \mapsto  \Psi'} = 
\frac{1}{\hbar^{2}} 
                   \Bigg\vert 
                   \int_{0}^{\tau} {}_I\langle \Psi | {\tilde H}_I |  \Psi' \rangle_I \, dt \, 
                   \Bigg\vert^{2},
\end{equation}   
where $|.\rangle_I$ denotes a ket in the interaction picture. In general,  ${\hat H}_I $ 
can be seen as a perturbation to $\hat H_0$ provided 
\begin{equation}                                                                         \label{cond}
\Delta( \Psi') := 
| \langle \Psi  | {\tilde H}_I | \Psi' \rangle | \, |\delta_{ \Psi'}|^{-1} \ll 1,
\end{equation}
where 
\begin{equation}                                                                         \label{det}
\delta_{ \Psi'}:= \langle \Psi  | \hat H_0 | \Psi \rangle - 
                  \langle \Psi'  | \hat H_0 | \Psi'\rangle .  
\end{equation}
For (\ref{in}), it is easy to show that 
\begin{equation}                                                                         \label{tp2}     
\mathcal P_{\Psi \mapsto  \Psi'} = 2 | \Delta( \Psi' )|^{2} 
                                [ 1 - \cos( \tfrac{1}{\hbar}\delta_{ \Psi'} \, \tau) ].
\end{equation}

The non-null terms in  $\langle \Psi | {\tilde H}_I |  \Psi' \rangle $ will give rise 
to finite transition probabilities that can be divided in two classes: 
energy conserving and non-energy conserving. 
Energy conserving transitions 
are those in which  a quantum of energy 
is simultaneously created in one oscillator and destroyed in the other. 
In the interaction Hamiltonian (\ref{in}),
terms responsible for these transitions are those proportional to 
${\tilde a}_j^{\dagger} {\tilde a}_k$ for $i\neq j$.  On the other hand, terms like 
${\hat a}_j {\hat a}_k$ and $ {\hat a}_j^{\dagger}{\hat a}_k^{\dagger}$ cause the 
net destruction or creation of two quanta of energy, either in one mode or one quantum in each mode. 
For the energy conserving transitions,  Eq.~(\ref{cond}) results in
\begin{equation}                                                                         \label{condce}
\Delta(\Psi'_\text{c})  = \Bigg| \frac{  {\eta}_{jk} \,   
                         \langle \mathfrak{n}_1, \mathfrak{n}_2 | 
                                            {\hat a}_j^{\dagger} {\hat a}_k  
                         | \Psi'_{\text c}\rangle                                          }
                     {  \omega_j - \omega_k   } \Bigg|,
\end{equation}
while for non-energy conserving ones it results in
\begin{equation}                                                                         \label{condnce1}
\Delta(\Psi'_{\text{nc}})  = \Bigg| \frac{  {\xi}_{jk} \,   
                         \langle \mathfrak{n}_1, \mathfrak{n}_2 | 
                                            {\hat a}_j {\hat a}_k  
                         | \Psi'_{\text{nc}}\rangle                                          }
                     {  \omega_j + \omega_k   } \Bigg|,  
\end{equation}
or 
\begin{equation}                                                                         \label{condnce2}
\Delta(\Psi'_{\text{nc}})  = \Bigg| \frac{  {\xi}^{\ast}_{jk} \,   
                         \langle \mathfrak{n}_1, \mathfrak{n}_2 | 
                                            {\hat a}_j^{\dagger} {\hat a}_k^{\dagger}  
                         | \Psi'_{\text{nc}}\rangle                                          }
                     {  \omega_j + \omega_k   } \Bigg|.  
\end{equation}

Now, it is worth noticing that if one approaches exact resonance ($\omega_j =\omega_k$), 
energy conserving transitions (\ref{condce}) will turn (\ref{tp2}) into 
\begin{equation}                                                                         \label{tpce}     
\mathcal P_{\Psi \mapsto \Psi'_{\text{c}}} \propto   \frac{\tau^2}{\hbar^{2}}, 
\end{equation}
representing a quadratically growth, while the non-conserving energy transitions 
(\ref{condnce1}) and (\ref{condnce2}) 
will still produce limited oscillations (\ref{tp2}) 
with very small amplitudes that are proportional to $1/(\omega_1+\omega_2)$. 
Consequently, the importance of the energy conservative terms soon supplants 
the importance of the non-conserving ones. 
Even in the case where one does not have exact resonance, 
the relative importance of the two kind of transitions will still 
favor the energy conserving ones, 
provided $|\omega_1-\omega_2|\ll |\omega_1+\omega_2|$. 
This is a much more relaxed condition compared to exact resonance. 
To see this, let us consider the common scenario where $ {\eta}_{jk}$ and ${\xi}_{jk}$ 
have the same order of magnitudes. Then, from (\ref{tp2}), (\ref{condce}), 
(\ref{condnce1}), and (\ref{condnce2}),  
it is easy to check that, apart from the limited oscillating functions, 
the relative importance of amplitudes is 
\begin{eqnarray}
\frac{\mathcal P_{\Psi \mapsto \Psi'_{\text c}}}
{\mathcal P_{\Psi \mapsto \Psi'_{\text nc}}}\approx\frac{|\omega_1+\omega_2|}{|\omega_1-\omega_2|}.
\end{eqnarray}
Based on these considerations, one can drop terms in the interaction picture Hamiltonian which 
oscillate with the sum of frequencies compared to the ones oscillating with the difference of frequencies.  
Alternatively, one can say that energy conserving transitions are the only ones to be kept in the Hamiltonian. 
This elimination of rapidly oscillating terms in the Hamiltonian is called RWA, regardless of whether one has exact or 
approximate resonance in the sense $|\omega_1-\omega_2|\ll |\omega_1+\omega_2|$. 

A word of caution is in order here. The RWA demands that the interaction part of the Hamiltonian $\hat H_I$, 
where the terms to be neglected lie, should be weak compared to the free part $\hat  H_0$. For (\ref{hamsimp}), 
this is guaranteed when $\eta_{jk},\xi_{jk}\ll\omega_1,\omega_2$. In this case, we were able to justify the 
RWA through first order perturbation theory. Otherwise, there is no reason for RWA to be valid. At higher orders, 
energy conserving and non-energy conserving terms mix in the perturbative series due to powers of $\hat H_I$.

Finally, in the RWA, and back to the Schr\"odinger picture, the system Hamiltonian  will be
\begin{equation}                                                                         \label{hamRWA1}
\hat H_{\rm eff} = 
\hbar (\omega_1 \hat a_1^{\dagger} \hat a_1 +\omega_2 \hat a_2^{\dagger} \hat a_2  + 
 {\eta}_{12} {\hat a}_1^{\dagger} {\hat a}_2+{\eta}_{12}^{\ast} {\hat a}_2^{\dagger} {\hat a}_1).   
\end{equation}
%

%%%%%%%%%%%%%%%%%%%%%%%%%%%%%%%%%%%%%%%%%%%%%%%%%%%%%%%%%%%%%%%%%%%%%%%%%%%%%%%%%%%%%%%%%
\section{ Matrix  \texorpdfstring{${\mathsf E}(t)$}{E(t)} and Auxiliary Functions } %%%%%
\label{ma}                                                            %%%%%%%%%%%%%%%%%%%
\renewcommand{\theequation}{B-\arabic{equation}}                      %%%%%%%%%%%%%%%%%%%
%%%%%%%%%%%%%%%%%%%%%%%%%%%%%%%%%%%%%%%%%%%%%%%%%%%%%%%%%%%%%%%%%%%%%%%%%%%%%%%%%%%%%%%%%
The simplectic matrix in (\ref{rwssimp2}) is written as 
\begin{equation}
{\mathsf E}( t ) = 
\begin{pmatrix}
 \mathbf C  & \mathbf S \\
-\mathbf S  & \mathbf C
\end{pmatrix},
\end{equation}
where we defined the matrices $\mathbf C = \cos(\mathbf{H_q} t) $ and 
$\mathbf S  = \sin(\mathbf{H_q}  t) $ with $\mathbf{H_q} $ given by Eq.~(\ref{hesseff2}).
Explicitly,

\begin{eqnarray}
{\mathbf C}_{11} &=& \chi^{-1}[ 1+ ( \chi -1 ) \cos ( \chi \tau )],                       \nonumber \\
{\mathbf C}_{12} &=& {\mathbf C}_{21} = \tfrac{2 {\mathsf S}_{m\alpha}}{\chi} 
                     \sin^2\left(\tfrac{ \chi \tau }{2} \right),                          \nonumber \\
{\mathbf C}_{13} &=& {\mathbf C}_{31} = \tfrac{2 {\mathsf S}_{m\beta}}{\chi} 
                      \sin^2\left(\tfrac{ \chi \tau }{2} \right),                         \nonumber \\  
{\mathbf C}_{22} &=& 
\tfrac{   {\mathsf S}_{m\beta}^2 }{ ( \chi - 1 ) } \cos\tau \,  + \,  
\tfrac{   {\mathsf S}_{m\alpha}^2}{\chi ( \chi - 1 )} [ (\chi-1) \, + \, \cos(\chi\tau) ],\nonumber \\
{\mathbf C}_{23} &=& {\mathbf C}_{32} =  
\tfrac{   {\mathsf S}_{m\alpha} {\mathsf S}_{m\beta} }{\chi ( \chi - 1 ) } 
          [ ( \chi - 1 ) - \chi\cos\tau \,  + \, \cos(\chi\tau)],                        \nonumber \\
{\mathbf C}_{33} &=&  
\tfrac{   {\mathsf S}_{m\alpha}^2 }{ ( \chi - 1 ) } \cos\tau \,  + \,  
\tfrac{   {\mathsf S}_{m\beta}^2}{\chi ( \chi - 1 )} [ (\chi-1) \, + \, \cos(\chi\tau) ],\nonumber 
\end{eqnarray}
and 

\begin{eqnarray}
{\mathbf S}_{11} &=& 
\tfrac{\chi-1}{\chi}  \sin(\chi\tau),                                                    \nonumber \\
{\mathbf S}_{12} &=&  {\mathbf S}_{21} =
-\tfrac{{\mathsf S}_{m\alpha} }{\chi}\sin( \chi \tau),                                   \nonumber \\
{\mathbf S}_{13} &=& {\mathbf S}_{31} = 
-\tfrac{{\mathsf S}_{m\beta} }{\chi}\sin( \chi \tau),                                    \nonumber \\
{\mathbf S}_{22} &=& 
\tfrac{   {\mathsf S}_{m\alpha}^2} {  ( \chi - 1 ) \chi} \sin(\chi\tau) + 
\tfrac{   {\mathsf S}_{m}^2} {  ( \chi - 1 )     } \sin \tau ,                     \nonumber \\
{\mathbf S}_{23} &=&  {\mathbf S}_{32} =  
\tfrac{ {\mathsf S}_{m\beta }{\mathsf S}_{m\alpha}}{ \chi ( \chi - 1 )} 
[\sin(\chi\tau)  - \chi \sin \tau ],                                                \nonumber \\
{\mathbf S}_{33} &=& 
\tfrac{   {\mathsf S}_{m\beta }^2} {\chi  ( \chi - 1 ) } \sin(\chi\tau) + 
\tfrac{   {\mathsf S}_{m\alpha}^2} {  ( \chi - 1 )     } \sin \tau ,                          
\end{eqnarray}
with $\tau := \epsilon t/4$, and 
$ \chi := {\mathsf S}_{m\alpha}^2 + {\mathsf S}_{m\beta}^2 + 1$.

%%%%%%%%%%%%%%%%%%%%%%%%%%%%%%%%%%%%%%%%%%%%%%%%%%%%%%%%%%%%%%%%%%%%%%%%%%%%%%%%%%%%%%%%%

%%%%%%%%%%%%%%%%%%%%%%%%%%%%%%%%%%%%%%%%%%%%%%%%%%%%%%%%%%%%%%%%%%%%%%%%%%%%%%%%%%%%%%%%%
\end{document}